\documentclass{ws-ijqi}

\usepackage{theorem}
\usepackage{enumerate}
\usepackage{amsmath}
\usepackage{amssymb}
\usepackage{graphicx}
\usepackage{epstopdf}
\usepackage{color}
\usepackage{euscript}
\usepackage{float}

\theorembodyfont{\rmfamily}
\newtheorem{Theorem}{\textit{Theorem} }
\newtheorem{Proposition}{Proposition}
\newtheorem{Definition}{\textit{Definition} }

\newtheorem{Lemma}{\textit{Lemma}}
\newtheorem{Proof}{\textit{Proof}}

\newcommand{\lv}{\left \vert}
\newcommand{\rv}{\right \vert}
\newcommand{\la}{\left \langle}
\newcommand{\ra}{\right \rangle}
\newcommand{\ket}[1]{\lv #1 \ra}
\newcommand{\bra}[1]{\la #1 \rv}
\newcommand{\braket}[2]{\langle #1 \vert #2 \rangle}
\newcommand{\ketbra}[2]{\lv #1 \rangle \langle #2 \rv}

\newcommand{\tr}{\mathrm{Tr}}

\begin{document}

\markboth{Yoshifumi Nakada and Mio Murao}
{Diagonal-unitary $2$-designs and their implementations by quantum circuits}

\catchline{}{}{}{}{}

\title{Diagonal-unitary $2$-designs and their implementations by quantum circuits  }

\author{Yoshifumi Nakata}

\address{Department of Physics, Graduate School of Science,
University of Tokyo, Tokyo, Japan\\
nakata@eve.phys.s.u-tokyo.ac.jp
\footnote{A current position is Institute for Theoretical Physics, Leibniz University Hannover, Germany. 
An email address yoshifumi.nakata@itp.uni-hannover.de is also available.}
}

\author{Mio Murao}

\address{Department of Physics, Graduate School of Science,
University of Tokyo, Tokyo, Japan\\
Institute for Nano Quantum Information Electronics,
 University of Tokyo, Tokyo, Japan
murao@phys.s.u-tokyo.ac.jp}

\maketitle

\begin{history}
\received{Day Month Year}
\revised{Day Month Year}
\end{history}

 \begin{abstract}
We study efficient generations of {\it random diagonal-unitary matrices}, an ensemble of unitary matrices diagonal in a given basis with randomly distributed phases for their eigenvalues.
Despite the simple algebraic structure, they cannot be achieved by quantum circuits composed of a few-qubit {\it diagonal} gates.
We introduce {\it diagonal-unitary $t$-designs}
and present 
two quantum circuits that implement diagonal-unitary $2$-designs with the computational basis in $N$-qubit systems.
One is composed of single-qubit diagonal gates and controlled-phase gates with randomized phases, which achieves an exact diagonal-unitary $2$-design after applying
the gates on all pairs of qubits. The number of required gates is $N(N-1)/2$.
If the controlled-Z gates are used instead of the controlled-phase gates, the circuit cannot achieve an exact $2$-design, but achieves
an $\epsilon$-approximate $2$-design by applying gates on {\it randomly} selected pairs of qubits. Due to the random choice of pairs, 
the circuit obtains extra randomness and the required number of gates is at most $O(N^2(N+\log1/\epsilon))$.
We also provide an application of the circuits, a protocol of generating an exact $2$-design of random states
by combining the circuits with a simple classical procedure requiring $O(N)$ random classical bits.
\end{abstract}



\section{Introduction}

Understanding typical properties of an ensemble is a useful methodology for analyzing quantum many-body systems.
When we investigate typical properties of all pure states in a Hilbert space, we use {\it random states}, an ensemble of pure states uniformly distributed in a given Hilbert space with respect to the unitarily invariant measure.
Random states have been intensively studied from many aspects, e.g., their entanglement\cite{L1978,P1993,FK1994,R1995,ZS2001,HLW2006}, a relation to an appearance of the Gibbs state in subsystems\cite{PSW2006,GLTZ2006,R2008,LPSW2009,RGE2012,VZ2012,MRA2011,BCHHKM2012}, a possible mechanism for black holes to have entropy\cite{HP2007}, and
quantum informational tasks\cite{L1997,RBSC2004,RRS2005,S2006,DCEL2009}.
Since random states are obtained by applying {\it random unitary matrices}\cite{M1990} to any pure state,
quantum circuit implementations of random unitary matrices are also studied\cite{DCEL2009,EWSLC2003,DLT2002,ODP2007,DOP2007,Z2008,HL2009,HL2009TPE,BH2010,BHH2012}.
It has been shown that the ensemble of unitary matrices
simulating up to the $t$-th order of statistical moments of random unitary matrices, referred to as a {\it unitary $t$-design},
can be approximately achieved by quantum circuits\cite{BHH2012}.
Random unitary matrices themselves have many utilities in quantum informational tasks\cite{DLT2002,BR2003,AMTW2000,HLSW2004,L2000,HHL2004,BHLSW2005,TDL2001,DBWR2010}.

{\it Phase-random states} are restricted random states, where only phases of the expansion coefficients in a given basis are randomly distributed\cite{NTM2012}.
They are originally introduced for investigating typical properties of time evolving states in isolated systems.
It has been shown that phase-random states with a separable basis and equal-amplitudes have extremely high entanglement on average\cite{NTM2012}.
This class of pure states have been studied in the context of locally maximally entangleable states\cite{KK2009,CBK2011,CKDV2013},
and in relation with mutually unbiased bases\cite{KR2005}.
They also reveal typical properties of states that appear in instantaneous quantum polynomial-time (IQP) circuits\cite{SB2009,BJS2010} and commuting circuits\cite{NV2012}, which are likely to have stronger computational power than classical computers even though they exploit only diagonal gates and a separable pure initial state.
However, it had not been yet clarified whether or not the corresponding phase-random states can be efficiently generated.


In this paper, we provide quantum circuits that efficiently generate phase-random states of $N$ qubits with the computational basis (tensor products of Pauli $Z$ basis).
To this end, we introduce {\it random diagonal-unitary matrices} and {\it diagonal-unitary $t$-designs}, which are
analogous to random unitary matrices and unitary $t$-designs, respectively.
In particular, we consider those with the computational basis.
One may think that such random diagonal-unitary matrices could be obtained by applying one-qubit diagonal gates with random phases. However, this is not the case 
since such random diagonal-unitary matrices generate large amount of entanglement even if an initial state is separable\cite{NTM2012},
but one-qubit gates cannot.
More precisely, only a diagonal-unitary $1$-design can be achieved by one-qubit diagonal gates.
A natural question is whether we can achieve a diagonal-unitary $t$-design ($t \geq 2$) by using two-qubit diagonal gates.

We show that {\it phase-random circuits} composed of two-qubit diagonal gates achieve diagonal-unitary $2$-designs.
Due to the commutativity of the gates, the circuit becomes stationary after applying the two-qubit gates on all pairs of qubits,
as long as the pairs are deterministically selected.
We first explicitly show that such a stationary circuit is a diagonal-unitary $2$-design 
if we use controlled-phase gates with two-valued random phases and single qubit phase gates with three-valued random phases. 
On the other hand, if we do not use random phases in a genuine two-qubit gate, e.g., 
the phases in each controlled-phase gate are fixed such that it becomes the controlled-$Z$ gate, 
the stationary circuit is not a diagonal-unitary $2$-design.
Hence, in the case of the phase-random circuit using the controlled-$Z$ gates, we need to introduce extra randomness to achieve a diagonal-unitary $2$-design.
We show that, if we apply two-qubit gates on {\it randomly} selected pairs of qubits, the phase-random circuit results in an {\it $\epsilon$-approximate} diagonal-unitary $2$-design
after applying at most $O(N^2(N+\log 1/\epsilon))$ gates.
In this case, the circuit does not become stationary after applying the gates on all pairs of qubits because of the random choices of the pairs.
These results show that random variables in genuine two-qubit gates enhance the ability of randomizing phases. 

We also provide an application of the phase-random circuit, i.e., a protocol for generating random states.
We show that an exact $2$-design of random states can be obtained 
by combining the phase-random circuit with a simple classical procedure requiring $O(N)$ random classical bits.
Since our protocol uses only diagonal gates in the quantum part of the protocol, it is simpler than previously known implementations particularly from an experimental point of view. 

This paper is organized as follows. In Sec.~\ref{Sec:RSPRS}, we give definitions of random unitary matrices, random diagonal-unitary matrices, and the corresponding designs. 
In Sec.~\ref{Sec:phase-random circuitSum}, we present our two main theorems on the phase-random circuits,
and the protocol generating an exact $2$-design of random states.
After explaining a sketch of the proofs of our theorems in Sec.~\ref{S:Sketch},
we present proofs of the first and second theorem in Sec.~\ref{Sec:phase-random circuitCP} and in Sec.~\ref{Sec:phase-random circuitCZ}, respectively.  
Finally, we summarize and present concluding remarks in Sec.~\ref{Sec:Sum}.

\section{Random matrices and Designs} \label{Sec:RSPRS}

We review definitions of random unitary matrices and random states.
Then, we present definitions of random diagonal-unitary matrices, phase-random states, and their corresponding designs.

In this paper, for simplicity, we denote by $\mathbb{E}$ any expectations over a probability distribution. When it is necessary, we explicitly write the space taken over for the expectation.

\begin{Definition}[Random unitary matrices and random states]
Let $\mathcal{U}(d)$ be the unitary group of degree $d$.
{\it Random unitary matrices} $\mathcal{U}_{\rm Haar}$ are the ensemble of unitary matrices uniformly distributed with respect to the Haar measure on $\mathcal{U}(d)$, $d\mu_{\rm Haar}$. {\it Random states} $\Upsilon_{\rm Haar}$ are defined by the ensemble of states $\{ U_{\mu} \ket{\Psi} \}_{U_\mu \in \mathcal{U}_{\rm Haar}}$ for any fixed state $\ket{\Psi} \in \mathcal{H}$.
\end{Definition}

Note that the distribution of random states is independent of the choice of the state $\ket{\Psi}$ since the Haar measure $d\mu_{\rm Haar}$ is 
unitarily invariant.

\begin{Definition}[Random diagonal-unitary matrices and phase-random states]
{\it Random diagonal-unitary matrices} with an orthonormal basis $\{ \ket{u_n} \},$ $\mathcal{U}_{\rm diag}(\{ \ket{u_n} \})$, are an ensemble of diagonal unitary matrices of the form $U_{\varphi} =  \sum_{n=1}^{d} e^{i \varphi_n} \ketbra{u_n}{u_n}$ where the phases $\varphi_n$ are uniformly distributed with respect to the normalized Lebesgue measure d$\varphi$ = d$\varphi_1 \cdots $d$\varphi_{d} / (2\pi)^{d} $ on $[0,2\pi)^d$.
For a given state $\ket{\Psi}$, 
{\it phase-random states} $\Upsilon_{\rm phase}(\{ |\braket{\Psi}{u_n} |, \ket{u_n } \})$ are an ensemble of states $\{ U_{\varphi} \ket{\Psi} \}_{U_{\varphi} \in \mathcal{U}_{\rm diag}(\{ \ket{u_n} \})}$.
\end{Definition}

In contrast to random unitary matrices, random diagonal-unitary matrices depend on 
the choice of the basis $\{ \ket{u_n} \}_n$. 
Consequently, phase-random states depend on  
the basis $\{ \ket{u_n} \}_n$ and the distribution of the amplitudes when the initial state is expanded in the basis, $\{ |\braket{\Psi}{u_n} | \}_n$.

A $t$-design of an ensemble is another finite ensemble that simulates up to the $t$-th order of moments of the original one on average. 
An $\epsilon$-approximate $t$-design is an ensemble that approximates the $t$-design, where $\epsilon$ is a degree of approximation.
In the case of designs of matrices, we evaluate the degree of approximation in terms of the diamond norm\cite{KSV2002}.
For a superoperator $\mathcal{E}$ on $\mathcal{H}$, the diamond norm is defined by
\begin{equation}
|\!| \mathcal{E} |\!|_{\diamond} := \sup_d \sup_{X \neq 0} \frac{|\!| (\mathcal{E} \otimes {\rm id}_d) X |\!|_1 }{ |\!|X|\!|_1}, \notag
\end{equation}
where ${\rm id}_d$ is the identity operator on another $d$-dimensional Hilbert space $\mathcal{H}^\prime$ and $X$ is any positive operator on $\mathcal{H} \otimes \mathcal{H^\prime}$.
To define an $\epsilon$-approximate $t$-design, let $\mathcal{V}$ be an ensemble of unitary matrices and
$\mathcal{E}_{\mathcal{V}}(\rho)$ be a superoperator such that 
\begin{align}
\mathcal{E}_{\mathcal{V}}^{(t)}(\rho) &:= \mathbb{E}_{\mathcal{V}} [ U_{\mu}^{\otimes t} \rho (U_{\mu}^{\dagger})^{\otimes t}], \label{Eq:superop}
\end{align}
for any states $\rho$ on a system consisting of $t N$ qubits.
Then, an $\epsilon$-approximate unitary $t$-design is defined as follows (see, e.g., Ref.~\refcite{HL2009}):

\begin{Definition}[{\it $\epsilon$-approximate unitary $t$-designs}] \label{Def:uni}
{\it Let $\mathcal{U}$ be random unitary matrices or random diagonal-unitary matrices.
An $\epsilon$-approximate $t$-design of $\mathcal{U}$, denoted by $\mathcal{U}^{(t,\epsilon)}$, is a finite ensemble of unitary matrices such that
\begin{equation*}
|\!|  \mathcal{E}^{(t)}_{\mathcal{U}} - \mathcal{E}^{(t)}_{\mathcal{U}^{(t,\epsilon)}}  |\! |_{\diamond} \leq \epsilon.
\end{equation*}}
\end{Definition}

Although there are several definitions of an $\epsilon$-approximate unitary $t$-design in terms of different measures of the distance,
they are shown to be all equivalent in the sense that, if $\mathcal{V}$ is an $\epsilon$-approximate unitary $t$-design in one of the definitions,
then it is also an $\epsilon'$-approximate unitary $t$-design in other definitions, where $\epsilon' = {\rm poly} (2^{tN}) \epsilon$ 
(see Ref.~\refcite{L2010} for details).

In the case of designs of states, we use a trace norm $|\!| X |\!|_1 = \tr |X|$ to evaluate the difference.

\begin{Definition}[{\it $\epsilon$-approximate state $t$-designs}]
{\it Let $\Upsilon$ be random states or phase-random states.
An $\epsilon$-approximate $t$-design of $\Upsilon$, denoted by $\Upsilon^{(t,\epsilon)}$, is a finite ensemble of states such that
\begin{equation*}
\biggl|\!\biggl|
\mathbb{E}_{\Upsilon^{(t,\epsilon)}} [ \ketbra{\psi_{\mu}}{\psi_{\mu}}^{\otimes t} ] 
- 
\mathbb{E}_{\Upsilon} [\ketbra{\psi_{\mu}}{\psi_{\mu}}^{\otimes t}]
 \biggr|\! \biggr|_{1} \leq \epsilon. 
\end{equation*}}
\end{Definition}

In case of $\epsilon=0$, the designs are called {\it exact} $t$-designs.
A $t$-design for random states is referred to as a {\it spherical $t$-design}\cite{DG1977,SZ1984,B2002}, a {\it complex-projective $t$-design}\cite{RBSC2004,AE2007} or a {\it quantum state $t$-design}\cite{HL2009}. In this paper, we call a $t$-design of random states a complex-projective $t$-design.
We also call a $t$-design of phase-random states a toric $t$-design
since the parameter space of phase-random states is a hypertorus.

We finally make a remark on the choice of phases in diagonal-unitary $t$-designs and toric $t$-designs.
It is sufficient to choose phases from a discrete set. For instance,
a set of unitary matrices $\Omega_t=\{ \sum_n e^{i \phi_n} \ketbra{u_n}{u_n} \}$,
where $\phi_n$ is randomly chosen from $\{\frac{2 \pi k}{t+1} \}_{k=0,1,\cdots, t}$, is a diagonal-unitary $t$-design with the basis $\{\ket{u_n}\}$ (see~\ref{Ap:DiscreteP} for details).
This simple fact means that we can use $(t+1)$-valued discrete random parameters instead of continuous random parameters.
However, even if the phases are randomly chosen from a discrete set, the implementation of such unitary matrices requires global 
randomizations of phases.
For implementing these matrices by quantum circuits, we need to decompose each unitary matrix into local
unitary operations, namely, one- and two-qubit gates, in an efficient way. This is the main concern of this paper.

\section{Main results and an application} \label{Sec:phase-random circuitSum}

\begin{figure}[tb]
\centering
  \includegraphics[width=50mm, clip]{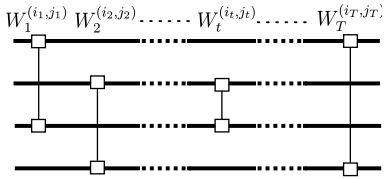}
  \caption{
A phase-random circuit. The vertical line denotes a two-qubit gate $W_t^{(i_t,j_t)}$ randomly selected from a diagonal two-qubit gate set $\mathcal{W}_{\rm diag}$.  
}
\label{Fig:phase-random circuit}
\end{figure}

In this section, we introduce phase-random circuits\cite{NTM2012} and provide our main results, i.e., phase-random circuits can achieve diagonal-unitary $2$-designs. 
We also present an application of phase-random circuit, a protocol generating an exact complex-projective $2$-design.

\subsection{Phase-random circuit}

We denote by $\ket{\bar{n}}$ the computational basis where $\bar{n}$ is a binary representation of $n-1$ ($n=1,\cdots,2^N$).
We investigate implementations of a diagonal-unitary $t$-design in the computational basis
$\mathcal{U}_{\rm diag}^{(t)}(\{\ket{\bar{n}} \} )$ by {\it phase-random circuits} presented in the previous work of the authors\cite{NTM2012}. For implementing a diagonal-unitary $t$-design in a general basis, it is sufficient to apply a unitary operation transforming the computational basis to the desired basis.

A phase-random circuit consists of $T$ diagonal two-qubit unitary gates shown in Fig.~\ref{Fig:phase-random circuit}.  For the $t$-th gate, we select two different numbers $(i_t,j_t)$ from $\{ 1, 2, \cdots , N\}$, as well as a two-qubit gate $W_t$ randomly from a given set of diagonal two-qubit gates $\mathcal{W}_{\rm diag}$.
We apply the two-qubit gate $W_t$ on the $i_t$-th and $j_t$-th qubits. An instance of the circuit is then specified by a set of parameters, $\mathcal{C}_T:=\{i_t,j_t,W_t\}_{t=1}^T$, and the unitary operation corresponding to the circuit is given by $U_T = W_T^{(i_T,j_T)} W_{T-1}^{(i_{T-1},j_{T-1})} \cdots W_{1}^{(i_1,j_1)}$, where $W_{t}^{(i_t,j_t)}$ non-trivially acts on the $i_t$-th and $j_t$-th qubits. Thus 
a phase-random circuit consisting of $T$ two-qubit gates is denoted by a set of the unitary operations $\{U_T\}_{\mathcal{C}_T}$.

\subsection{Main results} \label{SS:MR}

The main result of this paper is that,
if we choose an appropriate two-qubit diagonal gate set $\mathcal{W}_{\rm diag}$, the phase-random circuit achieves a diagonal-unitary $2$-design in the polynomial number of gates.  The necessary number of the gates depends on the choice of the gate set $\mathcal{W}_{\rm diag}$.

We consider two gate sets. First, we study the gate set given by 
\begin{equation*}
\mathcal{W}_{\rm diag}^{CP} = \{ \begin{pmatrix} 1 & 0 \\ 0 & e^{i \alpha} \end{pmatrix} \otimes \begin{pmatrix} 1 & 0 \\ 0 & e^{i \beta} \end{pmatrix}
\begin{pmatrix} 1 & 0 & 0 & 0 \\ 0 & 1 & 0 & 0 \\ 0 & 0 & 1 & 0 \\ 0 & 0 & 0 & e^{i \gamma} \end{pmatrix} \}_{ \alpha, \beta, \gamma \in [ 0,  2\pi)},
\end{equation*}
where all matrices are in the computational basis. 
In this case, we apply gates randomly drawn from $\mathcal{W}_{\rm diag}^{CP}$ on all the different pairs of qubits.
We refer to this phase-random circuit as a {\it CP phase-random circuit},
where the set of parameters is given by $\mathcal{C}_T^{CP} = \{\alpha_t, \beta_t, \gamma_t\}_{t=1}^T$.
The expectation is taken by
\begin{equation*}
\mathbb{E}_{\mathcal{C}_T^{CP}} [X] := \prod_{t=1}^T (\frac{1}{2 \pi} )^3 \int_{0}^{2 \pi} \! \! \! d\alpha_t \int_{0}^{2 \pi}\! \! \! d\beta_t \int_{0}^{2 \pi}\! \! \! d\gamma_t X.
\end{equation*}
For the CP phase-random circuit, we have the following theorem:

\begin{Theorem} \label{Thm:phase-random circuitCP}
{\it The CP phase-random circuit is a diagonal-unitary $2$-design in the computational basis.  
The number of the required gates is $\frac{N(N-1)}{2}$.}
\end{Theorem}

It is not necessary to choose the phases uniformly from $[0, 2\pi)$ and 
it suffices to choose the phases $\alpha$ and $\beta$ from $\{0, \frac{2 \pi}{3}, \frac{4 \pi}{3}\}$, 
and $\gamma$ from $\{0, \pi\}$ (see~\ref{App:DiscretePhasesCP} for details).
Thus, the CP phase-random circuit forms a finite ensemble and is a design.
Note that the controlled-phase gate with a random phase $\gamma \in \{0, \pi\}$ is the same as a probabilistic application of 
the controlled-$Z$ gate.
Since all gates in the CP phase-random circuit commute and are applied on deterministic choices of pairs of qubits, they can be applied simultaneously in a practical implementation.

For simplifying the implementation, one may think if we can get rid of random parameters in two-qubit gates.
Next, we deal with a gate set given by 
\begin{equation*}
\mathcal{W}_{\rm diag}^{CZ} = \{\begin{pmatrix} 1 & 0 \\ 0 & e^{i \alpha} \end{pmatrix} \otimes \begin{pmatrix} 1 & 0 \\ 0 & e^{i \beta} \end{pmatrix}
\begin{pmatrix} 1 & 0 & 0 & 0 \\ 0 & 1 & 0 & 0 \\ 0 & 0 & 1 & 0 \\ 0 & 0 & 0 & -1 \end{pmatrix} \}_{ \alpha, \beta \in [ 0, 2 \pi)}.
\end{equation*}
Similarly to the CP phase-random circuit, $\alpha, \beta$ can be chosen from $\{ 0, \frac{2\pi}{3}, \frac{4\pi}{3}  \}$ instead of $[0, 2\pi)$.
In this case, however, we cannot achieve an exact $2$-design even if we apply two-qubit gates randomly chosen from $\mathcal{W}_{\rm diag}^{CZ}$ on all pairs of qubits.
This fact is demonstrated in a three-qubit system as presented in the following.
When we apply two-qubit gates randomly chosen from $\mathcal{W}_{\rm diag}^{CZ}$ on every pair of three qubits, 
we obtain a unitary matrix given by
\begin{multline*}
U_3= {\rm diag} \biggl( 1, e^{i (\alpha_1 + \alpha_3) }, e^{i (\beta_1 + \alpha_2) }, - e^{i (\alpha_1 + \beta_1 + \alpha_2 + \alpha_3) },
e^{i (\beta_2 + \beta_3) }, -e^{i (\alpha_1 + \beta_2 + \alpha_3 + \beta_3) }, \\
 -e^{i (\beta_1 + \alpha_2 + \beta_2 + \beta_3 ) },
- e^{i (\alpha_1 + \beta_1 + \alpha_2 + \beta_2 +  \alpha_3 + \beta_3) } \biggr),
\end{multline*}
where $\alpha_i$ and $\beta_i$ are randomly chosen from $[0, 2\pi)$ or $\{0, \frac{2\pi}{3}, \frac{4\pi}{3} \}$ for every $i=1,2,3$.
Instead of checking if $\{ U_3 \}_{\alpha_i, \beta_i}$ is an exact diagonal-unitary $2$-design in terns of a superoperator $\mathcal{E}^{(2)}$,
we check an equivalent definition of an exact $2$-design\cite{HL2009} given by $\mathbb{E}_{\{\alpha_i, \beta_i\}} [U_3^{\otimes 2} \otimes   (U_3^{\dagger})^{\otimes 2}]  = 
\mathbb{E}_{\mathcal{U}_{\rm diag}} [U_{\mu}^{\otimes 2} \otimes   (U_{\mu}^{\dagger})^{\otimes 2}]$, where $\mathcal{U}_{\rm diag}$ denotes random diagonal-unitary matrices.
By a straightforward calculation, it can be shown that 
$\mathbb{E}_{\{\alpha_i, \beta_i\}} [U_3^{\otimes 2} \otimes   (U_3^{\dagger})^{\otimes 2}]$ contains $-1$ as an element,
but $\mathbb{E}_{\mathcal{U}} [U_{\mu}^{\otimes 2} \otimes   (U_{\mu}^{\dagger})^{\otimes 2}]$ does not.
That is,  $\{ U_3 \}_{\{\alpha_i, \beta_i\}}$ is not an exact diagonal-unitary $2$-design.
This difference comes from a fact that the controlled-Z gate does not have any parameters, resulting in a correlation between some elements of $U_3$ in a lower order, 
e.g., $(U_3)_{88} = (U_3)_{44} (U_3)_{55}$.

Thus we need to introduce extra randomness to achieve a diagonal-unitary $2$-design by using a gate set $\mathcal{W}_{\rm diag}^{CZ}$.
To this end, we choose $(i_t, j_t)$ randomly at each time, so that the set of parameters is given by $\mathcal{C}_T^{CZ} = \{i_t, j_t, \alpha_t, \beta_t\}_{t=1}^T$.
The expectation is given by 
\begin{equation}
\mathbb{E}_{\mathcal{C}_T^{CZ}} [X] := \prod_{t=1}^T \frac{2}{N(N-1)}\sum_{i_t \neq j_t} (\frac{1}{2 \pi} )^2 \int_{0}^{2 \pi} \! \! \! d\alpha_t \int_{0}^{2 \pi}\! \! \! d\beta_t X.
\end{equation}
Due to the random choice of $\{i_t,j_t\}$, the upper bound of $T$ can exceed $N(N-1)/2$ in spite of commutativity of all gates. 
We call the corresponding phase-random circuit a {\it CZ phase-random circuit}.
Note that parameters $\alpha$ and $\beta$ chosen from just $\{0, \frac{2 \pi}{3}, \frac{4 \pi}{3} \}$ guarantees the CZ phase-random circuit to form a finite ensemble.
The circuit achieves an approximate $2$-design, but not an exact $2$-design, as stated in the following theorem.

\begin{Theorem} \label{Thm:phase-random circuitCZ}
{\it The CZ phase-random circuit $\{U_T\}_{\mathcal{C}_T^{CZ}}$ consisting of $T$ two-qubit gates is an $\epsilon$-approximate diagonal-unitary $2$-design 
in the computational basis if $T \geq T_{conv}(\epsilon)$, where 
\begin{equation}
\frac{N}{2}  + \biggl(\frac{N^2}{4} + O(N) \biggr)\log (2\epsilon)^{-1}  \leq T_{conv}(\epsilon) 
\leq 7 N^3 \log2 + N^2 \log\epsilon^{-1} + O(N^2).  
\end{equation}
Therefore, the CZ phase-random circuit is an $\epsilon$-approximate diagonal-unitary $2$-design after applying at most
$O(N^2(N+\log \epsilon^{-1}))$ two-qubit gates.}
\end{Theorem}

In a practical sense, the CZ phase-random circuit has disadvantages compared to the CP phase-random circuit
since it cannot achieve an exact diagonal-unitary $2$-design.
Moreover, unlike the CP case,
the gates in the circuit cannot be applied simultaneously since the dynamics should be {\it stochastic} by choosing $i$ and $j$ randomly for each gate.
However, we present Theorem~\ref{Thm:phase-random circuitCZ} since the difference between the CZ and CP phase-random circuits show that a random parameter in the controlled-phase gate dramatically improves the ability to randomize, which we find theoretically interesting.

Finally, we emphasize that phase-random circuits are easily implementable in experiments
since they exploit only diagonal gates in the computational basis, which 
can be fault-tolerantly implemented in super- and semi-conductor systems\cite{ABDPST2009}.
Furthermore, in the case of the CP phase-random circuit,
all gates can be applied simultaneously, which significantly simplifies the experimental implementation.

\subsection{An application: a protocol for generating a complex-projective $2$-design} \label{SS:Appl}

Phase-random circuits generate toric $2$-designs exactly for the CP case and approximately for the CZ case. 
By combining phase-random circuits with an extra classical procedure, 
we can also obtain a complex-projective $2$-design, which has useful applications in quantum information processing\cite{L1997,RBSC2004,RRS2005,S2006,DCEL2009}. 
To show this, we first consider the difference between complex-projective and toric 2-designs.

The expectation of states for complex-projective $t$-designs is obtained from Schur's lemma\cite{SW1986} and is given by
\begin{align*}
\mathbb{E}_{\Upsilon_{\rm Haar}^{(t)}} [\ketbra{\psi}{\psi}^{\otimes t}]&=
\mathbb{E}_{\Upsilon_{\rm Haar}} [\ketbra{\psi}{\psi}^{\otimes t}]\\
&=\int_{\ket{\psi_{\mu}} \in \Upsilon_{\rm Haar}} \ketbra{\psi_{\mu}}{\psi_{\mu}}^{\otimes t} d\mu_{\rm Haar} \notag \\
&= \frac{1}{d_{\rm sym}}\Pi_{\rm sym}^{(t)}, 
\end{align*}
where $\Pi_{\rm sym}^{(t)}$ is a projector onto the symmetric subspace in $\mathcal{H}^{\otimes t}$ and $d_{\rm sym}$ is the dimension of the symmetric subspace. 
On the other hand, the expectation of states for toric $t$-designs is not a projector $\Pi_{\rm sym}^{(t)}$. For instance, in $N$-qubit systems, the expectation of states for the toric $2$-designs $\Upsilon^{(2)}_{\rm phase}(\{r_n, \ket{u_n} \}_{n} )$ is 
\begin{multline}
\mathbb{E}_{\Upsilon^{(2)}_{\rm phase}} [\ketbra{\psi}{\psi}^{\otimes 2}] 
= 
\sum_{n=1}^{2^N} r_n^4 \ketbra{u_n u_n}{u_n u_n} \\
+ 2\sum_{n > m} r_n^2 r_m^2 \frac{ \ket{u_n u_m} + \ket{u_m u_n}}{\sqrt{2}} \frac{ \bra{u_n u_m} + \bra{u_m u_n}}{\sqrt{2}}, \notag
\end{multline}
which is not proportional to the projector $\Pi_{\rm sym}^{(2)}$.

To close the gap, we consider a particular toric 2-design 
given by $\Upsilon^{(2)}_{\rm phase}(\{2^{-N/2}, \ket{\bar{n}} \} )$ where $\{\ket{\bar{n}} \}$ is the computational basis.
The expectation of the state over $\Upsilon^{(2)}_{\rm phase}(\{2^{-N/2}, \ket{\bar{n}} \})$ is given by 
\begin{align*}
\mathbb{E}_{\Upsilon^{(2)}_{\rm phase}} [\ketbra{\psi}{\psi}^{\otimes 2}] 
& \propto 
\sum_{n} \ketbra{\bar{n} \bar{n}}{\bar{n} \bar{n}} + 2\sum_{n > m} \frac{\ket{\bar{n} \bar{m}} + \ket{\bar{m} \bar{n}}}{\sqrt{2}} \frac{ \bra{\bar{n} \bar{m}} + \bra{\bar{m} \bar{n}}}{\sqrt{2}} \\
&= 2 \Pi_{\rm sym}^{(2)} -  \sum_{n} \ketbra{\bar{n} \bar{n}}{\bar{n} \bar{n}}. 
\end{align*}
This shows that a probabilistic mixture of $\Upsilon^{(2)}_{\rm phase}(\{2^{-N/2}, \ket{\bar{n}} \} )$ and a set of states $\{\ket{\bar{n}}\}$
forms a complex-projective 2-design.
Thus, the protocol to obtain a complex-projective 2-design using a phase-random circuit is given by
\begin{enumerate}
\item With probability $\frac{1}{2^N+1}$, choose a random bit $\bar{n}$ and generate a state $\ket{\bar{n}}$.
\item With probability $\frac{2^N}{2^N+1}$, perform the CP phase-random circuit with an initial state  $\ket{++ \cdots +}$.
\end{enumerate}

This protocol of generating an exact complex-projective $2$-design requires $O(N)$ random classical bits and $O(N^2)$ diagonal gates 
in the CP phase-random circuit. 
Below is the comparison with previously proposed protocols of generating exact and approximate complex-projective $2$-designs.

\begin{itemize}
\item An exact unitary 2-design using Clifford operations is known\cite{DLT2002}, which requires $O(N^8)$ bits and $O(N^2)$ quantum gates. 
In this protocol, a description of unitary matrices is classically calculated and is decomposed into one- and two-qubit unitary gates. 
Thus, for obtaining a sequence of states in a complex-projective $2$-design,
it is necessary to repeat calculating a gate decomposition and constructing the corresponding quantum circuit.
\item An $\epsilon$-approximate unitary 2-design is presented where the circuit is composed of one- and two-qubit Clifford gates and some gates are applied probabilistically\cite{DCEL2009}. The number of gates is $O(N \log 1/\epsilon)$ in their definition of the 2-design.
It has been pointed out that\cite{HL2009} it corresponds to $O(N(N+ \log 1/\epsilon))$ in Definition~\ref{Def:uni}.
\item Random circuits\cite{ODP2007,DOP2007} are known to converge to an $\epsilon$-approximate unitary 2-design
after applying $O(N(N+ \log 1/\epsilon))$ gates\cite{HL2009}.
In random circuits, all gates are randomly chosen from a set called a {\it 2-copy gapped} gate set, e.g., a set of the controlled-NOT gate and single qubit rotation gates.
\item A local random circuit composed of $O(N t^4( N + \log 1/\epsilon))$ gates is shown to form an $\epsilon$-approximate unitary $t$-design\cite{BHH2012}. The circuit is composed of random $SU(4)$ gates acting on nearest neighbor qubits.
\end{itemize}

In comparison with these results, the main advantage of our protocol is that it uses only {\it diagonal} gates.
Although the necessary number of gates in our protocol is the same as that of the previous results, the commutativity of diagonal gates enables us to
apply all gates simultaneously for implementation.
Moreover, diagonal gates greatly simplify a realization of the circuit in experiments.
The CP phase-random circuit requires only single-qubit rotations around the $Z$-axis and the random application of the controlled-$Z$ gate,
which can be fault-tolerantly implemented in super- and semi-conductor systems\cite{ABDPST2009}.
Thus, our protocol provides a way to generate a complex-projective 2-design by currently achievable technology.

\section{Sketch of the proofs} \label{S:Sketch}
In order to prove Theorem~\ref{Thm:phase-random circuitCP} and~\ref{Thm:phase-random circuitCZ}, 
we analyze how two-qubit diagonal gates in phase-random circuits transform an initial state.
Since our goal is to show the phase-random circuits are diagonal-unitary $2$-designs, we consider an initial state $\rho_0$
on a $2N$-qubit system (see Definition.~\ref{Def:uni} and Eq.~\eqref{Eq:superop}).
  
The technique is similar to that used for investigating the convergence of random circuits\cite{ODP2007,DOP2007}.
We denote the state after applying $T$ two-qubit gates by $\rho_T= (U_T)^{\otimes 2} \rho_0 (U_T^{\dagger})^{\otimes 2}$ and 
expand it in the Pauli basis to investigate the evolution of each coefficient.
We denote by $\mathbf{p}$ and $\mathbf{q}$ vectors corresponding to the subscripts of the Pauli basis of two $N$-qubit systems $(p_1,\cdots, p_N)$ and $(q_1,\cdots, q_N)$, where $p_i, q_i \in \{ 0,x,y,z \}$, respectively. Then, the state $\rho_T$ is expressed by
\begin{equation}
\rho_T
=2^{-N} \sum_{\mathbf{p}, \mathbf{q}} \xi_T (\mathbf{p},\mathbf{q}) \sigma_{\mathbf{p}} \otimes \sigma_{\mathbf{q}}, \notag
\end{equation}
where $\sigma_{\mathbf{p}}:=\sigma_{p_1} \otimes \cdots \otimes \sigma_{p_N}$ and
$\sigma_{\mathbf{q}}:=\sigma_{q_1} \otimes \cdots \otimes \sigma_{q_N}$ are tensor products of the Pauli operators.
Similarly, we consider a state $\rho_{\varphi} =(U_{\varphi})^{\otimes 2} \rho_0 (U_{\varphi}^{\dagger})^{\otimes 2}$, 
where $U_{\varphi}$ is an element of a diagonal-unitary $2$-design $\mathcal{U}_{\rm diag}^{(2)}$, and expand $\rho_{\varphi}$ in the Pauli basis
\begin{equation}
\rho_{\varphi}
=2^{-N} \sum_{\mathbf{p}, \mathbf{q}} \xi_{\varphi} (\mathbf{p},\mathbf{q}) \sigma_{\mathbf{p}} \otimes \sigma_{\mathbf{q}}. \notag
\end{equation}

To simplify the following investigation, we introduce a notation.
Since the way how $\xi_T(\mathbf{p},\mathbf{q})$ is transformed depends on $\sigma_{\mathbf{p}}$ and $\sigma_{\mathbf{q}}$,
it is convenient to define subsets in $\{1,\cdots, N \}$ that specify the locations of $\sigma_w$ ($w=0,x,y,z$) in $\sigma_{\mathbf{p}}$ and $\sigma_{\mathbf{q}}$;
\begin{align*}
&\Gamma^{(+)} (\mathbf{p}, \mathbf{q}) := \{ i \in \{1,\cdots, N \} | p_i=q_i \in \{x, y \} \}, \\
&\Gamma^{(-)} (\mathbf{p}, \mathbf{q}) := \{ i \in \{1,\cdots, N \} | p_i=\bar{q}_i \in \{x, y \} \}, \\
&\Lambda^{(+)} (\mathbf{p}, \mathbf{q}) := \{ i \in \{1,\cdots, N \} | p_i=q_i \in \{0, z \} \},\\
&\Lambda^{(-)} (\mathbf{p}, \mathbf{q}) := \{ i \in \{1,\cdots, N \} | p_i=\bar{q}_i \in \{0, z \} \},
\end{align*}
where the bar  sign of $\bar{q}_i$ represents to take a self-inverse `flip' map defined by $\bar{0}=z$ and $\bar{x}=y$.
We denote the number of elements in each subset by  the corresponding small letters, e.g.,
$\gamma^{(\pm)}(\mathbf{p},\mathbf{q})$ ($\lambda^{(\pm)}(\mathbf{p},\mathbf{q})$) is the number of elements in $\Gamma^{(\pm)}(\mathbf{p},\mathbf{q})$ ($\Lambda^{(\pm)}(\mathbf{p},\mathbf{q})$).
We also denote the union of 
$\Gamma^{(+)}(\mathbf{p},\mathbf{q})$ and $\Gamma^{(-)}(\mathbf{p},\mathbf{q})$ ($\Lambda^{(+)}(\mathbf{p},\mathbf{q})$ and $\Lambda^{(-)}(\mathbf{p},\mathbf{q})$) by $\Gamma(\mathbf{p},\mathbf{q})$ ($\Lambda(\mathbf{p},\mathbf{q})$). 
Similarly, the number of elements in $\Gamma(\mathbf{p},\mathbf{q})$ ($\Lambda(\mathbf{p},\mathbf{q})$) is
denoted by $\gamma(\mathbf{p},\mathbf{q})$ ($\lambda(\mathbf{p},\mathbf{q})$). By definition, 
$\gamma(\mathbf{p},\mathbf{q})=\gamma^{(+)}(\mathbf{p},\mathbf{q})+\gamma^{(-)}(\mathbf{p},\mathbf{q})$ and
$\lambda(\mathbf{p},\mathbf{q})=\lambda^{(+)}(\mathbf{p},\mathbf{q})+\lambda^{(-)}(\mathbf{p},\mathbf{q})$.
We introduce $\Gamma^{(-)}_{even}(\mathbf{p},\mathbf{q})$ and $\Gamma^{(-)}_{odd}(\mathbf{p},\mathbf{q})$ as subsets of $\Gamma^{(-)}(\mathbf{p},\mathbf{q})$ of which the number of elements is even and odd, respectively.

We also define a function $f_S(\mathbf{p})$ of $\mathbf{p}$, where $S=\{i_1, \cdots, i_s \}$ is a subset of $\{1, \cdots, N \}$,
such as $f_S(\mathbf{p})= (p_1, \cdots, \bar{p}_{i_1}, \cdots, \bar{p}_{i_s}, \cdots, p_N)$. 
That is, the function $f_S$ flips all elements of $\mathbf{p}$ in the set $S$.
For instance, $f_{1,3} (y,x,0) = (\bar{y},x ,\bar{0}) = (x,x,z)$.
We denote $(f_S(\mathbf{p}), f_S(\mathbf{q}))$ simply by $f_S(\mathbf{p}, \mathbf{q})$. 
Finally, we define
$S_{even}(\mathbf{p}, \mathbf{q})$ and $S_{odd}(\mathbf{p}, \mathbf{q})$ by
$S_{even}(\mathbf{p}, \mathbf{q})=\{f_s(\mathbf{p}, \mathbf{q}) \text{ where } s \subset \Lambda(\mathbf{p},\mathbf{q}) \cup \Gamma^{(+)}(\mathbf{p},\mathbf{q}) \cup \Gamma^{(-)}_{even}(\mathbf{p},\mathbf{q})    \}$ and
$S_{odd}(\mathbf{p}, \mathbf{q})=\{f_s(\mathbf{p}, \mathbf{q}) \text{ where } s \subset \Lambda(\mathbf{p},\mathbf{q}) \cup \Gamma^{(+)}(\mathbf{p},\mathbf{q}) \cup \Gamma^{(-)}_{odd}(\mathbf{p},\mathbf{q})    \}$.
For simplicity, we often omit the part $(\mathbf{p}, \mathbf{q})$ in equations when there is no ambiguity.

In terms of the expansion coefficients $\xi_T (\mathbf{p},\mathbf{q})$ and $\xi_{\varphi} (\mathbf{p},\mathbf{q})$, 
our goal is to show that for any initial state $\rho_0$,
\begin{equation}
\forall (\mathbf{p},\mathbf{q}), \biggl|\mathbb{E}_{\mathcal{C}_T}[\xi_T (\mathbf{p},\mathbf{q}) ] - \mathbb{E}_{\mathcal{U}_{\rm diag}^{(2)}}[\xi_{\varphi} (\mathbf{p},\mathbf{q}) ] \biggr| \leq \frac{\epsilon}{2^{7N}}, \label{Eq:StateError}
\end{equation}
after sufficiently large $T$. 
In~\ref{App:NecSuff}, we show that
if Eq.~\eqref{Eq:StateError} holds for any initial state, the phase-random circuit is an $\epsilon$-approximate diagonal-unitary $2$-design.
$\mathbb{E}_{\mathcal{U}_{\rm diag}^{(2)}}[\xi_{\varphi} (\mathbf{p},\mathbf{q})]$
for an initial state $\rho_0 =2^{-N} \sum_{\mathbf{p}, \mathbf{q}} \xi_0 (\mathbf{p},\mathbf{q}) \sigma_{\mathbf{p}} \otimes \sigma_{\mathbf{q}}$
is obtained by a tedious but straightforward calculation.
For $(\mathbf{p},\mathbf{q})$ such that $\lambda(\mathbf{p},\mathbf{q}) = N$, 
$\mathbb{E}_{\mathcal{U}_{\rm diag}^{(2)}} [\xi_{\varphi}(\mathbf{p},\mathbf{q})] = \xi_{0}(\mathbf{p},\mathbf{q})$.
For $(\mathbf{p},\mathbf{q})$ satisfying $\gamma(\mathbf{p},\mathbf{q}) + \lambda(\mathbf{p},\mathbf{q}) =N$ ($\lambda(\mathbf{p},\mathbf{q}) \neq N$), 
\begin{align}
\mathbb{E}_{\mathcal{U}_{\rm diag}^{(2)}} &[\xi_{\varphi}(\mathbf{p},\mathbf{q})] 
=
2^{-N}\biggl[ \sum_{S_{even}(\mathbf{p}, \mathbf{q})} -  
\sum_{S_{odd}(\mathbf{p}, \mathbf{q})} \biggr] \xi_0(\mathbf{p}', \mathbf{q}'), \label{Eq:Goal}
\end{align}
where the summation is taken over all $(\mathbf{p}', \mathbf{q}') \in  S_{even (odd)}(\mathbf{p}, \mathbf{q})$.
Otherwise, $\mathbb{E}_{\mathcal{U}_{\rm diag}^{(2)}} [\xi_{\varphi} (\mathbf{p},\mathbf{q})] =0$.

In Section~\ref{Sec:phase-random circuitCP}, we present that the CP phase-random circuit achieves Eq.~\eqref{Eq:StateError} for $\epsilon=0$, which gives the proof of Theorem~\ref{Thm:phase-random circuitCP}.
In Section~\ref{Sec:phase-random circuitCZ}, we show that Eq.~\eqref{Eq:StateError} holds for the CZ phase-random circuit after at most applying $O(N^2(N+ \log1/\epsilon))$ two-qubit diagonal gates, implying Theorem~\ref{Thm:phase-random circuitCZ}.

\section{CP phase-random circuit} \label{Sec:phase-random circuitCP}

We present a proof of Theorem~\ref{Thm:phase-random circuitCP}. 
To do so, we follow the transformation of the expectation of the expansion coefficient $\mathbb{E}_{\mathcal{C}_T^{CP}}[\xi_{T+1}(\mathbf{p},\mathbf{q})]$ by the CP phase-random circuit.
For simplicity, hereafter we omit the subscript $\mathcal{C}_T^{CP}$ for the expectation value. 

By applying $W_{T+1}(\alpha_{T+1}, \beta_{T+1}, \gamma_{T+1})$ on a pair of qubits specified by two numbers $(i, j)$, the expectation of the coefficients changes to
\begin{equation*}
\mathbb{E} [\xi_{T+1}(\mathbf{p},\mathbf{q})] 
= 2^{-2N} \sum_{(\mathbf{p}',\mathbf{q}')} 
G_{ij}(\mathbf{p},\mathbf{q};\mathbf{p}',\mathbf{q}') \mathbb{E} [\xi_{T}(\mathbf{p}',\mathbf{q}')] ,
\end{equation*}
where the matrix $G_{ij}(\mathbf{p},\mathbf{q};\mathbf{p}',\mathbf{q}')$ is given in~\ref{Ap:G} by 
\begin{equation*}
G_{ij}(\mathbf{p},\mathbf{q};\mathbf{p}',\mathbf{q}')= 
\mathbb{E}[\tr \sigma_{\mathbf{p}} W_{T+1} \sigma_{\mathbf{p}'} W_{T+1}^{\dagger} \tr \sigma_{\mathbf{q}} W_{T+1} \sigma_{\mathbf{q}'} W_{T+1}^{\dagger} ].  
\end{equation*}
Then, we obtain
\begin{align}
\mathbb{E}[\xi_{T+1}(\mathbf{p},\mathbf{q})]
=
\begin{cases}
\mathbb{E}[\xi_T(\mathbf{p},\mathbf{q})] & \text{if } i,j \in \Lambda(\mathbf{p},\mathbf{q}),\\
\frac{1}{4}\biggl( \mathbb{E}[\xi_T(\mathbf{p},\mathbf{q})] \pm \mathbb{E}[\xi_T(f_{i}(\mathbf{p},\mathbf{q}))]\\
\hspace{5mm} +\mathbb{E}[\xi_T(f_{j}(\mathbf{p},\mathbf{q}))] \pm \mathbb{E}[\xi_T(f_{ij}(\mathbf{p},\mathbf{q}))]\biggr)  & \text{if } i \in \Gamma^{(\pm)}(\mathbf{p},\mathbf{q}) \text{, } j \in \Lambda(\mathbf{p},\mathbf{q}),\\
\frac{1}{4} \biggl(
\mathbb{E}[\xi_T(\mathbf{p},\mathbf{q})] + v \mathbb{E}[\xi_T( f_j (\mathbf{p},\mathbf{q}) )] \\
\hspace{5mm} +u \mathbb{E}[\xi_T( f_i(\mathbf{p},\mathbf{q}))] +uv  \mathbb{E}[\xi_T( f_{ij}(\mathbf{p},\mathbf{q}))]\biggr)  & \text{if } i \in \Gamma^{(u)}(\mathbf{p},\mathbf{q}) \text{, } j \in \Gamma^{(v)}(\mathbf{p},\mathbf{q}), \\
0 & \text{otherwise},
\end{cases} \label{Eq:EvoCP}
\end{align}
where $u$ and $v$ are $\pm1$.

Note that a set of $\Lambda(\mathbf{p}, \mathbf{q})$ and $\Gamma(\mathbf{p}, \mathbf{q})$ are both invariant under the transformation
since the transformation is composed of a function $f_S(\mathbf{p}, \mathbf{q})$ which flips $x$ ($y$) to $y$ ($x$) and $0$ ($z$) to $z$ ($0$).
This implies that for a given $\Lambda' \subset \{1,\cdots, N\}$ and $\Gamma' \subset \{1,\cdots, N\}$,
only the coefficients $\mathbb{E}[\xi_{T}(\mathbf{p},\mathbf{q})]$ for $(\mathbf{p}, \mathbf{q})$ such that
$\Lambda(\mathbf{p}, \mathbf{q}) = \Lambda'$ and $\Gamma(\mathbf{p}, \mathbf{q}) = \Gamma'$ mix up each other by the transformation.

From the last case in Eq.~\eqref{Eq:EvoCP}, it is clear that for any pairs of indices $(\mathbf{p},\mathbf{q})$ satisfying  $\gamma(\mathbf{p},\mathbf{q}) + \lambda(\mathbf{p},\mathbf{q}) <N$, we have $\mathbb{E}[\xi_{T}(\mathbf{p},\mathbf{q})]=0$ for any $T$ after $i  \notin \Lambda(\mathbf{p},\mathbf{q}) \cup \Gamma(\mathbf{p},\mathbf{q})$ is chosen.
The other cases show that when one of $(i,j)$ is in $\Gamma(\mathbf{p},\mathbf{q})$, we take the uniform average of the `flipped' terms.
In particular, when $i \in \Gamma^{(-)}(\mathbf{p},\mathbf{q})$ is selected, the term acquires a negative sign.
Hence, after all pairs of qubits are drawn, the expectation of the state becomes an average of all flipped terms with appropriate negative signs.
Then, we have the following proposition.

\begin{Proposition} \label{Prop:CPA}
{\it After applying $W_{ij}$ to all combinations of $i$ and $j$, which requires $T_{CP}=\frac{N(N-1)}{2}$ two-qubit gates, the coefficient converges to
\begin{equation}
\mathbb{E}[\xi_{T_{CP}}(\mathbf{p},\mathbf{q})]=2^{-N}\biggl[ \sum_{S_{even}(\mathbf{p}, \mathbf{q})} -  
\sum_{S_{odd}(\mathbf{p}, \mathbf{q})} \biggr] \xi_0(\mathbf{p}', \mathbf{q}'), \label{Eq:Owari}
\end{equation}
where the summation is taken over all $(\mathbf{p}', \mathbf{q}') \in  S_{even (odd)}(\mathbf{p}, \mathbf{q})$.}
\end{Proposition}

Since Eq.~\eqref{Eq:Owari} is the same as Eq.~\eqref{Eq:Goal}, 
we obtain that for any initial states and for all $(\mathbf{p}, \mathbf{q})$,  
\begin{equation}
\mathbb{E}[\xi_{T_{CP}}(\mathbf{p},\mathbf{q})] = \mathbb{E}_{\mathcal{U}^{(2)}_{\rm diag} }[\xi_{\varphi}(\mathbf{p},\mathbf{q}) ]. \notag
\end{equation}
This concludes the proof of Theorem~\ref{Thm:phase-random circuitCP}.

\section{CZ phase-random circuit} \label{Sec:phase-random circuitCZ}

Theorem~\ref{Thm:phase-random circuitCZ} is shown by proving the following two lemmas. 
The first lemma guarantees that the state transformed by the CZ phase-random circuit for any initial state converges to the corresponding toric $2$-design.
The second lemma states that the convergence time $T_{conv}(\epsilon)$ defined in Theorem~\ref{Thm:phase-random circuitCZ} scales as the cube of the system size $N$.

\begin{Lemma} \label{Lem:phase-random circuitCZ}
{\it For any initial state and $\forall (\mathbf{p}, \mathbf{q})$,
\begin{equation*}
\lim_{T \rightarrow \infty} \biggl|\mathbb{E}_{\mathcal{C}_T^{CZ}}[\xi_T (\mathbf{p},\mathbf{q}) ] - \mathbb{E}_{\mathcal{U}_{\rm diag}^{(2)}}[\xi_{\varphi} (\mathbf{p},\mathbf{q}) ] \biggr|=0.  
\end{equation*}}
\end{Lemma}

\begin{Lemma} \label{Lem:MixingTime}
{\it For any initial state, the convergence time $T_{conv}(\epsilon)$ satisfies
\begin{equation*}
\frac{N}{2}  + \biggl(\frac{N^2}{4} + O(N) \biggr)\log (2\epsilon)^{-1}   \leq T_{conv}(\epsilon) \leq 7 N^3 \log2 + N^2 \log\epsilon^{-1} + O(N^2).  
\end{equation*}
Therefore, $T_{conv}(\epsilon)$ is at most $O(N^3) + O(N^2) \log \epsilon^{-1}$.}
\end{Lemma}
We present the proof of Lemma~\ref{Lem:phase-random circuitCZ} in Subsection~\ref{SS:Lem1} and the proof of Lemma~\ref{Lem:MixingTime} in Subsection~\ref{SS:lemma2}.

\subsection{Convergence of the distribution} \label{SS:Lem1}

We consider a transformation of a state by the CZ phase-random circuit.
Similarly to the CP case, we obtain (see~\ref{Ap:G})
\begin{align}
\mathbb{E}[\xi_{T+1}(\mathbf{p},\mathbf{q})]
=
\begin{cases}
\mathbb{E}[\xi_T(\mathbf{p},\mathbf{q})] & \text{if } i,j \in \Lambda(\mathbf{p},\mathbf{q}),\\
\frac{1}{2}( \mathbb{E}[\xi_T(f_j(\mathbf{p},\mathbf{q}))] \pm \mathbb{E}[\xi_T(f_{ij}(\mathbf{p},\mathbf{q}))])  & \text{if } i \in \Gamma^{(\pm)}(\mathbf{p},\mathbf{q}) \text{, } j \in \Lambda(\mathbf{p},\mathbf{q}),\\
\frac{1}{4} (
\mathbb{E}[\xi_T(\mathbf{p},\mathbf{q})] + v \mathbb{E}[\xi_T( f_j (\mathbf{p},\mathbf{q}) )] \\
\hspace{5mm} +
u \mathbb{E}[\xi_T( f_i(\mathbf{p},\mathbf{q}))] + u v \mathbb{E}[\xi_T( f_{ij}(\mathbf{p},\mathbf{q}))]
)  & \text{if } i \in \Gamma^{(u)}(\mathbf{p},\mathbf{q}) \text{, } j \in \Gamma^{(v)}(\mathbf{p},\mathbf{q}), \\
0 & \text{otherwise}.
\end{cases} \label{Eq:Evo1}
\end{align}
where $u, v \in \{+, - \}$.
The transformation is different from that of the CP phase-random circuits given by Eq.~\eqref{Eq:EvoCP} only when
$i \in \Gamma^{(\pm)}(\mathbf{p},\mathbf{q})$ and $j \in \Lambda(\mathbf{p},\mathbf{q})$.
This prevents the circuit from randomizing the corresponding phases. Thus, we have to introduce stochastic transformations by choosing $(i,j)$ randomly 
in order to achieve a diagonal-unitary $2$-design.

\begin{figure}[tb]
\centering
  \includegraphics[width=45mm, clip]{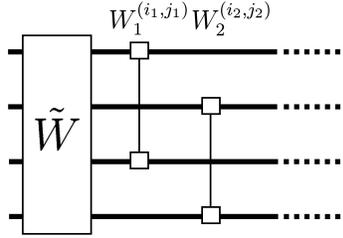}
  \caption{
A modified phase-random circuit. A $N$-qubit unitary operation $\tilde{W}$ is applied in advance of the phase-random circuit.
The unitary operation $\tilde{W}$ is composed of two-qubits diagonal gates acting on neighboring qubits, $\tilde{W}=W_{N,N-1} \cdots W_{3,2} W_{2,1}$.
}
\label{Fig:Mphase-random circuit}
\end{figure}

To simplify the investigation, we introduce a {\it modified phase-random circuit}.
Similarly to the CP case, for any pairs of indices $(\mathbf{p},\mathbf{q})$ satisfying  $\gamma(\mathbf{p},\mathbf{q}) + \lambda(\mathbf{p},\mathbf{q}) <N$, we have 
$\mathbb{E}[\xi_{T}(\mathbf{p},\mathbf{q})]=0$ for any $T$ once after $i  \notin \Lambda(\mathbf{p},\mathbf{q}) \cup \Gamma(\mathbf{p},\mathbf{q})$ is chosen.
In order to avoid the complication by dealing with such $(\mathbf{p},\mathbf{q})$,
we first apply a two-qubit gate $W_{i,j} (\alpha, \beta)$ on all neighboring qubits $(2k-1,2k)$.
We denote this unitary operations by $\tilde{W} = W_{N,N-1}W_{N-2,N-3} \cdots  W_{2,1}$. (see Fig.~\ref{Fig:Mphase-random circuit}) 
When $N$ is odd,  we define $\tilde{W}$ by  $W_{N,N-1}W_{N-1,N-2} \cdots  W_{2,1}$.
The number of two-qubit gates required to perform $\tilde{W}$ is $T_{\tilde{W}} = \lceil N/2 \rceil$ where $\lceil n \rceil$ is the smallest integer
larger or equal to $n$. Note that $\tilde{W}$ is composed of commuting gates and is deterministic. 
Hence, they can be applied simultaneously.

In analogy with Proposition~\ref{Prop:CPA}, we obtain the following proposition.

\begin{Proposition}  \label{Prop:Before}
{\it If
$\gamma(\mathbf{p},\mathbf{q}) + \lambda(\mathbf{p},\mathbf{q}) <N$, $\mathbb{E}[\xi_{T_{\tilde{W}}}(\mathbf{p},\mathbf{q})]=0$.
If $(\mathbf{p},\mathbf{q})$ satisfies
$\gamma(\mathbf{p},\mathbf{q}) + \lambda(\mathbf{p},\mathbf{q}) = N$,
\begin{equation*}
\mathbb{E}[\xi_{T_{\tilde{W}}}(\mathbf{p},\mathbf{q})]
=2^{-\gamma} \biggl[\sum_{s \subset \Gamma^{(-)}_{even}} - \sum_{s \subset \Gamma^{(-)}_{odd}}  \biggr]
\sum_{s' \subset \Gamma^{(+)}}
\xi_0(f_{s \cup s'} \circ f_{\tilde{\Lambda}} (\mathbf{p},\mathbf{q})),  
\end{equation*}
where $(\mathbf{p},\mathbf{q})$ is omitted for simplicity and $\tilde{\Lambda}(\mathbf{p}, \mathbf{q})$ is a set of 
$i \in \Lambda(\mathbf{p}, \mathbf{q})$ such that $i$ is paired with an element of $\Gamma(\mathbf{p}, \mathbf{q})$ in $\tilde{W}$.}
\end{Proposition}

After applying $\tilde{W}$ consisting of
$T_{\tilde{W}}$ two-qubit gates, additional $T^\prime$ two-qubit diagonal gates randomly selected from $\mathcal{W}_{\rm diag}^{CP}$ are applied
on a randomly chosen pair of qubits. The following proposition shows how $\mathbb{E}[\xi_{T}(\mathbf{p},\mathbf{q})]$ relates to $\mathbb{E}[\xi_{T+1}(\mathbf{p},\mathbf{q})]$ for $T = T_{\tilde{W}} + T^\prime $.

\begin{Proposition} \label{Prop:after}
{\it For $T > T_{\tilde{W}}$, 
\begin{equation}
\mathbb{E}[\xi_{T+1}(\mathbf{p},\mathbf{q})] = 
\sum_{(\mathbf{p}',\mathbf{q}')}\mathcal{G}(\mathbf{p},\mathbf{q};\mathbf{p}',\mathbf{q}')\mathbb{E}[\xi_{T}(\mathbf{p}',\mathbf{q}')], \label{Eq:PropAfter}
\end{equation}
where $\mathcal{G}(\mathbf{p},\mathbf{q};\mathbf{p}',\mathbf{q}')$ is equal to
$\frac{\lambda ( \lambda -1 ) + \gamma ( \gamma -1 )}{N(N-1)}$ if $(\mathbf{p}',\mathbf{q}')=(\mathbf{p},\mathbf{q})$, $\frac{2 \gamma }{N(N-1)}$ if $(\mathbf{p}',\mathbf{q}')=f_i(\mathbf{p},\mathbf{q}) \text{ for } i \in \Lambda(\mathbf{p},\mathbf{q})$ and $0$ otherwise.}
\end{Proposition}
 
\begin{Proof}
The equation~\eqref{Eq:PropAfter} is equivalent to
\begin{equation}
\mathbb{E}[\xi_{T+1}(\mathbf{p},\mathbf{q})] = 
\frac{\lambda ( \lambda -1 ) + \gamma ( \gamma -1 )}{N(N-1)}\mathbb{E}[\xi_{T}(\mathbf{p},\mathbf{q})]
+\frac{2 \gamma }{N(N-1)} \sum_{i \in \Lambda(\mathbf{p},\mathbf{q})} \mathbb{E}[\xi_{T}(f_i(\mathbf{p},\mathbf{q}))]. \label{Eq:Kore}
\end{equation}
We show Eq.~\eqref{Eq:Kore} recursively.

Firstly, we show that
$\mathbb{E}[\xi_{T_{\tilde{W}}+1}(\mathbf{p},\mathbf{q})]=
\mathbb{E}[\xi_{T_{\tilde{W}}}(f_i(\mathbf{p},\mathbf{q}))]$ when $i \in \Lambda(\mathbf{p},\mathbf{q}) \text{ and } j \in \Gamma(\mathbf{p},\mathbf{q})$, and $\mathbb{E}[\xi_{T_{\tilde{W}}+1}(\mathbf{p},\mathbf{q})]=\mathbb{E}[\xi_{T_{\tilde{W}}}(\mathbf{p},\mathbf{q})]$ otherwise. 
This is derived by using Eq.~\eqref{Eq:Evo1} and Proposition~\ref{Prop:Before} in the following way.
When $i, j \in \Lambda(\mathbf{p},\mathbf{q})$, 
Eq.~\eqref{Eq:Evo1} implies $\mathbb{E}[\xi_{T_{\tilde{W}}+1}(\mathbf{p},\mathbf{q})]=\mathbb{E}[\xi_{T_{\tilde{W}}}(\mathbf{p},\mathbf{q})]$.
When $i \in \Lambda(\mathbf{p},\mathbf{q})$ and $j \in \Gamma^{(\pm)}(\mathbf{p},\mathbf{q})$, 
$\mathbb{E}[\xi_{T_{\tilde{W}}+1}(\mathbf{p},\mathbf{q})]$ is calculated to be
\begin{align*}
\mathbb{E}[\xi_{T_{\tilde{W}}+1}(\mathbf{p},\mathbf{q})] &= 
\frac{1}{2} \mathbb{E}[\xi_{T_{\tilde{W}}}(f_i(\mathbf{p},\mathbf{q}))] \pm \frac{1}{2} \mathbb{E}[\xi_{T_{\tilde{W}}}(f_{ij}(\mathbf{p},\mathbf{q}))]\\
&=\mathbb{E}[\xi_{T_{\tilde{W}}}(f_i(\mathbf{p},\mathbf{q}))],
\end{align*}
where the second line of the equation is obtained from the relation
\begin{equation}
\forall  j \in \Gamma^{(\pm)}(\mathbf{p},\mathbf{q}), \mathbb{E}[\xi_{T_{\tilde{W}}} (f_j(\mathbf{p},\mathbf{q}))]=
\pm \mathbb{E}[\xi_{T_{\tilde{W}}}(\mathbf{p},\mathbf{q})]. \label{Eq:Import}
\end{equation}
When $i, j \in \Gamma(\mathbf{p},\mathbf{q})$, a direct calculation shows that $\mathbb{E}[\xi_{T_{\tilde{W}}+1}(\mathbf{p},\mathbf{q})]=\mathbb{E}[\xi_{T_{\tilde{W}}}(\mathbf{p},\mathbf{q})]$ by using Eq.~\eqref{Eq:Import}.
Thus, we obtain the statement.

Since the probability that $j \in \Gamma(\mathbf{p},\mathbf{q})$ is chosen for a fixed $i \in \Lambda(\mathbf{p},\mathbf{q})$
is given by $\gamma/\binom{N}{2}$, where $\binom{N}{2}$ is a binomial coefficient, Eq.~\eqref{Eq:Kore} is shown for $T=T_{\tilde{W}}$.
Moreover, $\mathbb{E}[\xi_{T_{\tilde{W}}+1}(\mathbf{p},\mathbf{q})]$ also satisfies 
$\mathbb{E}[\xi_{T_{\tilde{W}}+1} (f_i(\mathbf{p},\mathbf{q}))]=
\pm \mathbb{E}[\xi_{T_{\tilde{W}}+1}(\mathbf{p},\mathbf{q})]$ for $i \in \Gamma^{(\pm)}(\mathbf{p},\mathbf{q})$, so that
Eq.~\eqref{Eq:Kore} is recursively obtained.
\begin{flushright}$\blacksquare$\end{flushright}
\end{Proof}

Proposition~\ref{Prop:after} implies that, for $T>T_{\tilde{W}}$, $\mathbb{E}[\xi_{T}(\mathbf{p},\mathbf{q})]$ is given by a convex sum of 
$\mathbb{E}[\xi_{T_{\tilde{W}}}(f_s(\mathbf{p},\mathbf{q}))]$ where $s$ is a subset of $\Lambda(\mathbf{p},\mathbf{q})$.
We define subsets $L^{(\pm)}, R^{(\pm)} \subsetneq \{1,\cdots, N\}$ where each subset is mutually exclusive.  For such subsets, transformation of $\mathbb{E}[\xi_{T}(\mathbf{p},\mathbf{q})]$ is closed in $\Sigma(L^{(\pm)},R^{(\pm)}):=\{(\mathbf{p},\mathbf{q}) | \Lambda^{(\pm)}(\mathbf{p},\mathbf{q})=L^{(\pm)},  \Gamma^{(\pm)}(\mathbf{p},\mathbf{q})=R^{(\pm)}   \}$.
In Proposition~\ref{Prop:Stat}, we consider transformation in $\Sigma(L^{(\pm)}, R^{(\pm)})$ and derive
the stationary distribution $\mathbb{E}[\xi_{\infty}(\mathbf{p},\mathbf{q})]:=\lim_{T \rightarrow \infty} \mathbb{E}[\xi_{T}(\mathbf{p},\mathbf{q})]$ for $(\mathbf{p},\mathbf{q}) \in \Sigma(L^{(\pm)}, R^{(\pm)})$.

\begin{Proposition} \label{Prop:Stat}
{\it 
Let $L^{(\pm)}$ and $R^{(\pm)}$ be appropriate subsets of $\{1,\cdots, N\}$ where each subset is mutually exclusive
and $L:=L^{(+)} \cup L^{(-)} \neq \{1, \cdots, N\}$.
For $ (\mathbf{p},\mathbf{q}) \in \Sigma(L^{(\pm)},R^{(\pm)})$, 
the stationary distribution $\mathbb{E}[\xi_{\infty}(\mathbf{p},\mathbf{q})]$ is uniform in 
$\Sigma(L^{(\pm)},R^{(\pm)})$, that is,
\begin{equation}
\mathbb{E}[\xi_{\infty}(\mathbf{p},\mathbf{q})]
=
\frac{1}{2^{l}} \sum_{(\mathbf{p}',\mathbf{q}') \in \Sigma(L^{(\pm)},R^{(\pm)})} 
\mathbb{E}[\xi_{T_{\tilde{W}}}(\mathbf{p}',\mathbf{q}')], \label{Eq:Stat1}
\end{equation}
where $l$ is the number of elements of $L$, which is also equal to $\lambda(\mathbf{p},\mathbf{q})$.
Moreover, for any $(\mathbf{p},\mathbf{q})$, we obtain
\begin{equation}
\mathbb{E}[\xi_{\infty}(\mathbf{p},\mathbf{q})]
=\mathbb{E}_{\mathcal{U}^{(2)}_{\rm diag}} [\xi_{\varphi}(\mathbf{p},\mathbf{q}) ] . \label{Eq:Trivial}
\end{equation} }
\end{Proposition}

In order to prove Proposition~\ref{Prop:Stat}, we use the Perron-Frobenius theorem\cite{HJ1985} for {\it irreducible} and {\it aperiodic} non-negative matrices $M$. Irreducibility is the property that for all $i$ and $j$ there exists a natural number $n$ such that $(M^n)_{ij}>0$ and
aperiodicity is the property that $M_{ii}>0$ for all $i$. A non-negative matrix is such that $M_{ij} \geq 0$ for all $i$ and $j$. The Perron-Frobenius theorem is given by the following statement.
\begin{Theorem}[Perron-Frobenius theorem\cite{HJ1985}]
{\it If a non-negative matrix $M$ is irreducible and aperiodic, the maximum eigenvalue $\lambda >0 $ is uniquely determined.
Let $\ket{\lambda}$ be the eigenvector corresponding to the maximum eigenvalue, then,
$\lim_{n \rightarrow \infty} (\frac{1}{\lambda} M)^n = \ket{\lambda}\bra{\lambda}$.}
\end{Theorem}

In addition to irreducibility and the aperiodicity, when a non-negative matrix $M$ is {\it bistochastic}, that is, $\sum_i M_{ij}=\sum_j M_{ij}=1 $, 
the maximum eigenvalue $\lambda$ is known to be equal to $1$. 
By applying these facts, we prove Proposition~\ref{Prop:Stat} as follows.
 
\begin{Proof}
In the case of $L=\{1, \cdots, N\}$, it is straightforward to prove from Eq.~\eqref{Eq:Evo1}.
When $L\neq \{1, \cdots, N\}$, we obtain Eq.~\eqref{Eq:Stat1} by applying the Perron-Frobenius theorem to the matrix $\mathcal{G}$ in $\Sigma(L^{(\pm)}, R^{(\pm)})$. 
If we restrict the matrix $\mathcal{G}$ to $\Sigma(L^{(\pm)}, R^{(\pm)})$, 
$\mathcal{G}$ is an irreducible, aperiodic and bistochastic non-negative matrix in $\Sigma(L^{(\pm)}, R^{(\pm)})$.
Hence, the Perron-Frobenius theorem guarantees that there exists a unique stationary distribution in $\Sigma(L^{(\pm)}, R^{(\pm)})$.
Since the evolution governed by $\mathcal{G}$ in $\Sigma(L^{(\pm)}, R^{(\pm)})$ is uniform, the stationary distribution is also uniform, resulting that, $\forall (\mathbf{p},\mathbf{q}) \in \Sigma(L^{(\pm)},R^{(\pm)})$,
\begin{equation*} 	
\mathbb{E}[\xi_{\infty}(\mathbf{p},\mathbf{q})]=
\frac{1}{2^l} \sum_{(\mathbf{p}',\mathbf{q}') \in \Sigma (L^{(\pm)}, R^{(\pm)})} 
\mathbb{E}[\xi_{T_{\tilde{W}}}(\mathbf{p}',\mathbf{q}')].   
\end{equation*}
Thus, we obtain Eq.~\eqref{Eq:Stat1}.

Next, we show Eq.~\eqref{Eq:Trivial}.
For $(\mathbf{p},\mathbf{q})$ satisfying $\lambda(\mathbf{p},\mathbf{q}) + \gamma(\mathbf{p},\mathbf{q}) < N$,  $\mathbb{E}[\xi_{\infty}(\mathbf{p},\mathbf{q})]=0$ since all $\mathbb{E}[\xi_{T_{\tilde{W}}}(\mathbf{p},\mathbf{q})]$ in the right hand side of Eq.~\eqref{Eq:Stat1} are zero as shown in Proposition~\ref{Prop:Before}.
Since Eq.~\eqref{Eq:Goal} implies $\mathbb{E}[\xi_{\mathcal{U}^{(2)}_{\rm diag}}(\mathbf{p},\mathbf{q})]=0$ for such 
$(\mathbf{p},\mathbf{q})$, we obtain $\mathbb{E}[\xi_{T_{\infty}}(\mathbf{p},\mathbf{q})]=\mathbb{E}[\xi_{\mathcal{U}^{(2)}_{\rm diag}	}(\mathbf{p},\mathbf{q})]$.
When $(\mathbf{p},\mathbf{q})$ satisfies $\lambda(\mathbf{p},\mathbf{q}) + \gamma(\mathbf{p},\mathbf{q}) = N$,
we substitute $\mathbb{E}[\xi_{T_{\tilde{W}}}(\mathbf{p},\mathbf{q})]$ given in Proposition~\ref{Prop:Before} into 
Eq.~\eqref{Eq:Stat1}, and obtain 
\begin{equation*}
\mathbb{E}[\xi_{\infty}(\mathbf{p},\mathbf{q})]=2^{-N}\biggl[ \sum_{S_{even}(\mathbf{p}, \mathbf{q})} -  
\sum_{S_{odd}(\mathbf{p}, \mathbf{q})} \biggr] \xi_0(\mathbf{p}', \mathbf{q}'), 
\end{equation*}
which is equal to $\mathbb{E}_{\mathcal{U}^{(2)}_{\rm diag}} [\xi_{\varphi}(\mathbf{p},\mathbf{q}) ]$.
\begin{flushright}$\blacksquare$\end{flushright}
\end{Proof}

\subsection{Convergence time for the phase-random circuits} \label{SS:lemma2}

In this subsection, we investigate the convergence time $T_{\rm conv}(\epsilon)$ defined by the condition that $\forall T \geq T_{conv}(\epsilon)$,
\begin{equation}
|\!| \mathcal{E}^{(2)}_{\{ U_T \}_{C_T^{CZ}}} - \mathcal{E}^{(2)}_{\mathcal{U}_{\rm diag}^{(2)}} |\!|_{\diamond} \leq  \epsilon, \label{Eq:ConvTime1norm}
\end{equation}
where $\mathcal{E}^{(2)}_{\mathcal{V}}$ for an ensemble of unitary matrices $\mathcal{V}$ is a superoperator defined by Eq.~\eqref{Eq:superop}.
A sufficient condition for Eq.~\eqref{Eq:ConvTime1norm} to hold is
\begin{equation}
\forall (\mathbf{p},\mathbf{q}), \biggl| \mathbb{E}_{C_T^{CZ}}[ \xi_T (\mathbf{p},\mathbf{q})  ]-\mathbb{E}_{C_{\infty}^{CZ}}[ \xi_{\infty} (\mathbf{p},\mathbf{q})  ] \biggr| 
\leq \frac{\epsilon}{2^{7N}}. \label{Eq:CoeffError}
\end{equation}
Similarly, we obtain a necessary condition for Eq.~\eqref{Eq:ConvTime1norm} 
by evaluating a lower bound of the diamond norm:
\begin{equation}
\forall (\mathbf{p},\mathbf{q}), \biggl| \mathbb{E}_{C_T^{CZ}}[ \xi_T (\mathbf{p},\mathbf{q})  ]-\mathbb{E}_{C_{\infty}^{CZ}}[ \xi_{\infty} (\mathbf{p},\mathbf{q})  ] \biggr| \leq \epsilon. \label{Eq:CoeffErrorLower}
\end{equation}
See~\ref{App:NecSuff} for details of derivations of these conditions.
Note that $\mathbb{E}_{C_{\infty}^{CZ}}[ \xi_{\infty} (\mathbf{p},\mathbf{q})] = \mathbb{E}_{\mathcal{U}_{\rm diag}^{(2)}}[ \xi_{\varphi} (\mathbf{p},\mathbf{q})] $ from Lemma~\ref{Lem:phase-random circuitCZ}.

We derive an upper and a lower bound on $T_{\rm conv}(\epsilon)$ by using Eq.~\eqref{Eq:CoeffError} and Eq.~\eqref{Eq:CoeffErrorLower}, respectively, and  prove Lemma~\ref{Lem:MixingTime}.
For $(\mathbf{p}, \mathbf{q})$ satisfying $\lambda(\mathbf{p},\mathbf{q}) + \gamma(\mathbf{p},\mathbf{q}) < N$,
$T_{\rm conv}(\epsilon) \leq T_{\tilde{W}} = \lceil N/2 \rceil$ since $\mathbb{E}[\xi_{T}(\mathbf{p},\mathbf{q})]=0$ for $T \geq T_{\tilde{W}}$, and, for $({\bf p}, {\bf q})$ satisfying $\lambda(\mathbf{p}, \mathbf{q}) = N$,
$T_{\rm conv}(\epsilon) =0 $ since $\mathbb{E}[\xi_{T}(\mathbf{p},\mathbf{q})]=\xi_{0}(\mathbf{p},\mathbf{q})$.
In the following, we consider only $(\mathbf{p}, \mathbf{q})$ such that 
$\gamma(\mathbf{p}, \mathbf{q}) + \lambda(\mathbf{p}, \mathbf{q}) = N$ and $\lambda(\mathbf{p}, \mathbf{q}) \neq N$. 

In order to show Lemma~\ref{Lem:MixingTime}, we use a technique of a Markov chain\cite{LPW2009}. 
We provide a brief introduction of a Markov chain in~\ref{Ap:Markov}. 
We map the transformation of $\mathbb{E}[\xi_{T}(\mathbf{p},\mathbf{q})]$ into a Markov chain and give a lower and upper bounds of the convergence time.

\subsubsection{Map to a Markov chain} \label{SSS:Map}
We present a map from the transformation of $\mathbb{E}[\xi_{\infty}(\mathbf{p},\mathbf{q})]$ by the (modified) CZ phase-random circuit to a Markov chain.
As shown in Proposition~\ref{Prop:after}, the transformation of $\mathbb{E}[\xi_{T}(\mathbf{p},\mathbf{q})]$ is restricted to $\Sigma(L^{(\pm)}, R^{(\pm)})$
and $\mathcal{G}[(\mathbf{p},\mathbf{q});(\mathbf{p}',\mathbf{q}')]$ satisfies the Markov property.
Moreover, it is observed from Proposition~\ref{Prop:after} that $\mathcal{G}[(\mathbf{p},\mathbf{q});(\mathbf{p}',\mathbf{q}')]$ is equivalent to the transition matrix of a random walk on a $l$-dimensional hypercube where each vertex is given by $(\mathbf{p}, \mathbf{q}) \in \Sigma(L^{(\pm)}, R^{(\pm)})$. Note that $\mathcal{G}$ is irreducible and aperiodic in $\Sigma(L^{(\pm)}, R^{(\pm)})$.
However, $\mathbb{E}[\xi_{T}(\mathbf{p},\mathbf{q})]$ cannot be regarded as a probability distribution since they are not necessarily non-negative. Instead, we define a probability distribution in the following way.

We set the initial probability distribution $\{\Pi_0 (\mathbf{p}, \mathbf{q}) \}$ of the Markov chain $\mathcal{M}(L^{(\pm)}, R^{(\pm)})$ to be 
\begin{equation*}
\Pi_0 (\mathbf{p}, \mathbf{q}) := \frac{\mathbb{E}[\xi_{T_{\tilde{W}}}(\mathbf{p},\mathbf{q})] - \Pi_{\min}(L^{(\pm)}, R^{(\pm)}) }{\Pi_{\rm sum}(L^{(\pm)}, R^{(\pm)})},  
\end{equation*}
where
\begin{equation*}
\Pi_{\rm min}(L^{(\pm)}, R^{(\pm)}) := \min_{(\mathbf{p},\mathbf{q}) \in \Sigma(L^{(\pm)}, R^{(\pm)})} \mathbb{E}[\xi_{T_{\tilde{W}}}(\mathbf{p},\mathbf{q})],  
\end{equation*}
and
\begin{equation*}
\Pi_{\rm sum}(L^{(\pm)}, R^{(\pm)}) := 
\sum_{(\mathbf{p},\mathbf{q}) \in \Sigma(L^{(\pm)}, R^{(\pm)})} \biggl( \mathbb{E}[\xi_{T_{\tilde{W}}}(\mathbf{p},\mathbf{q})] - \Pi_{\rm min}(L^{(\pm)}, R^{(\pm)}) \biggr).  
\end{equation*}
When there is no ambiguity, we omit $(L^{(\pm)}, R^{(\pm)})$ for $\Pi_{\rm min}$ and $\Pi_{\rm sum}$.
Then the probability distribution $\{\Pi_1 (\mathbf{p}, \mathbf{q}) \}$ is calculated to be
\begin{align*}
\Pi_1 (\mathbf{p}, \mathbf{q}) &= \sum_{(\mathbf{p}', \mathbf{q}') \in \Sigma(L^{(\pm)}, R^{(\pm)})}
\mathcal{G} [(\mathbf{p},\mathbf{q}); (\mathbf{p}',\mathbf{q}')]\Pi_0 (\mathbf{p}', \mathbf{q}') \\
&=\frac{1}{\Pi_{\rm sum}}( \mathbb{E}[\xi_{T_{\tilde{W}}+1}(\mathbf{p},\mathbf{q})] - \Pi_{\min}),
\end{align*}
where we use Proposition~\ref{Prop:after} and the fact that the matrix $\mathcal{G}$ is bistochastic.
Repeating this, the probability distribution after $k$ steps is
\begin{equation}
\Pi_k (\mathbf{p}, \mathbf{q})=\frac{1}{\Pi_{\rm sum}}( \mathbb{E}[\xi_{T_{\tilde{W}}+k}(\mathbf{p},\mathbf{q})] - \Pi_{\min}). \label{Eq:Pik}
\end{equation}
Thus we can define a Markov chain $\mathcal{M}(L^{(\pm)}, R^{(\pm)})$ on a $l$-dimensional hypercube with transition matrix $\mathcal{G}[(\mathbf{p},\mathbf{q});(\mathbf{p}',\mathbf{q}')]$ and probability distribution $\Pi_k (\mathbf{p}, \mathbf{q})$.

Note that Eq.~\eqref{Eq:Pik} leads to $\forall (\mathbf{p},\mathbf{q}) \in \Sigma(L^{(\pm)}, R^{(\pm)})$
\begin{equation*}
\biggl| \Pi_k (\mathbf{p}, \mathbf{q}) - \Pi_{\infty} (\mathbf{p}, \mathbf{q}) \biggr| 
=\frac{1}{\Pi_{\rm sum}} \biggl| \mathbb{E}[\xi_{T_{\tilde{W}}+k}(\mathbf{p},\mathbf{q})] - \mathbb{E}[\xi_{\infty}(\mathbf{p},\mathbf{q})]\biggr|,  
\end{equation*}
where $\Pi_{\infty} (\mathbf{p}, \mathbf{q})$ is the stationary distribution of the Markov chain $\mathcal{M}(L^{(\pm)}, R^{(\pm)})$.
This implies that
if the Markov chain $\mathcal{M}(L^{(\pm)}, R^{(\pm)})$ converges with an error $\epsilon/\Pi_{\rm sum}$,
$\mathbb{E}[\xi_{T_{\tilde{W}}}(\mathbf{p},\mathbf{q})]$ converges with an error $\epsilon$.
Hence $T_{\rm conv}(\epsilon) =T_{\rm mix} (\epsilon/\Pi_{\rm sum})$ where $T_{\rm mix}$ is the mixing time of the Markov chain defined in~\ref{Ap:Markov}.

The mixing time of the Markov chain on the hypercube depends on two factors.
One is the dimension of the hypercube $l$ and another is the probability that no change happens for a Markov chain, which is called a {\it staying probability}. Obviously, a larger $l$ and a smaller staying probability result in a longer mixing time.
For the Markov chain $\mathcal{M}(L^{(\pm)}, R^{(\pm)})$, the maximum of the dimension and the minimum of the staying probability are achieved 
for $l=N-1$. If the Markov chain $\mathcal{M}(L^{(\pm)}, R^{(\pm)})$ with $l=N-1$ converge with an error $\epsilon/\Pi_{\rm sum}$,
the other Markov chains with $l \neq N-1$ converges with an error less than $\epsilon/\Pi_{\rm sum}$. 
Thus, hereafter, we consider only the Markov chain $\mathcal{M}(L^{(\pm)}, R^{(\pm)})$ with $l=N-1$, which we denote by $\mathcal{M}$.

\begin{figure}[tb]
\centering
  \includegraphics[width=45mm, clip]{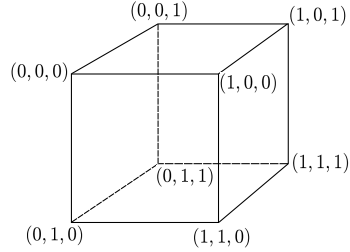}
  \caption{
A $3$-dimensional cube. A random walk on the cube is equivalent to a Markov chain $\mathcal{M}$ with $N=3$.}
\label{Fig:HC}
\end{figure}

Finally, we simplify the notation of the Markov chain $\mathcal{M}$. Since it is equivalent to a Markov chain on a $(N-1)$-dimensional hypercube
with a transition matrix $\mathcal{G}$, we label each vertex by a binary number $\vec{i}=(i_1,\cdots, i_{N-1}) \in \{0,1 \}^{N-1}$, not by $(\mathbf{p},\mathbf{q})$.
If we identify $(\mathbf{p}_0, \mathbf{q}_0)$ with $\vec{i_0}= 0 \cdots 0$, a new label for $f_k(\mathbf{p}_0, \mathbf{q}_0)$  is given by
$0 \cdots 0 {1}0 \cdots 0$ where only the $k$-th digit of the binary representation is $1$. The vertices $\vec{i}$ and $\vec{j}$ are connected if and only if $H(\vec{i},\vec{j})=1$ where $H(\vec{i},\vec{j})=\sum_k |i_k-j_k |$ (see Fig.~\ref{Fig:HC}).
In this new labeling, the transition matrix $\mathcal{G}[(\mathbf{p}, \mathbf{q});(\mathbf{p}', \mathbf{q}')]$ is simplified to 
\begin{equation}
\mathcal{P} (\vec{i},\vec{j})
=
\begin{cases}
1-\frac{2}{N} & \text{ if } \vec{i}=\vec{j}, \\
\frac{2}{N(N-1)} & \text{ if } H(\vec{i},\vec{j})=1, \\
0 & \text{ otherwise}.\label{Eq:TranP}
\end{cases}  
\end{equation}
 
\subsubsection{Lower bound on the mixing time of a Markov chain $\mathcal{M}$} \label{SSS:Lower}

The Markov chain on a $(N-1)$-dimensional hypercube with a staying probability $1/2$ has been well studied. For such a Markov chain, the transition matrix is given by $I/2 + \mathcal{P}'/2$ where $\mathcal{P}'$ has a matrix element $\frac{1}{N-1}$ for $H(\vec{i},\vec{j})=1$, and $0$ otherwise.
All eigenvalues of the transition matrix are known to be $\{1-\frac{k}{N-1}\}_{k=0,\cdots,N-1}$\cite{LPW2009}.
On the other hand, it is observed from Eq.~\eqref{Eq:TranP} that $\mathcal{P} (\vec{i},\vec{j})= (1-\frac{2}{N})I + \frac{2}{N}\mathcal{P}'$.
Hence, the eigenvalues of $\mathcal{P}(\vec{i},\vec{j})$ are given by $\{1-\frac{4k}{N(N-1)}\}_{k=0,\cdots,N-1}$,
which leads to $\max_{i=2,3,\cdots}|\lambda_i| = 1-\frac{4}{N(N-1)}$.
Using Eq.~\eqref{Eq:RelaxMarkov} in~\ref{Ap:Markov}, the mixing time $T_{\rm mix}(\epsilon')$ is bounded from below by
\begin{equation*}
(\frac{N(N-1)}{4} -1)  \log \frac{1}{2\epsilon'} \leq T_{\rm mix}(\epsilon').  
\end{equation*}

\subsubsection{Upper bound of the mixing time of a Markov chain $\mathcal{M}$} \label{SSS:Coupling}

In order to derive an upper bound on the mixing time, we slightly modify the transition matrix $\mathcal{P} (\vec{i},\vec{j})$
by changing the staying probability from $1-\frac{2}{N}$ to $1-\frac{1}{N-1}$.
A new transition matrix $\tilde{\mathcal{P}}(\vec{i},\vec{j})$ is given by 
\begin{equation*}
\tilde{\mathcal{P}} (\vec{i},\vec{j})
=
\begin{cases}
1-\frac{1}{N-1} & \text{ if } \vec{i}=\vec{j}, \\
\frac{1}{(N-1)^2} & \text{ if } H(\vec{i},\vec{j})=1, \\
0 & \text{ otherwise}. 
\end{cases}
\end{equation*}
We denote by $\tilde{\mathcal{M}}$ a new Markov chain with a transition matrix  $\tilde{\mathcal{P}}$.
Since the staying probability of $\mathcal{P}$ is smaller than $\tilde{\mathcal{P}}$, the mixing time of $\tilde{\mathcal{M}}$ provides
an upper bound on that of $\mathcal{M}$.

We investigate the mixing time of the Markov chain $\tilde{\mathcal{M}}$ by the {\it coupling method} (see~\ref{Ap:Markov}).
For constructing a coupling, we interpret $\tilde{\mathcal{M}}$ as follows.
At a step $t=k_0 + m(N-1)$ ($k_0=1, \cdots, N-1$ and $m=0,1, \cdots$), choose $j \in \{i_1, \cdots, i_{N-1}  \}$ at random and, if $j =i_{k_0}$, flip $i_{k_0}$, otherwise, do nothing. 
Based on this, we define a coupling $(X_k, Y_k)$. A state in $X_k$ ($Y_k$) is a bit sequence $x_1 \cdots x_{N-1}$ ($y_1 \cdots y_{N-1}$) where $x_j$ ($y_j$) $\in \{0,1 \}$.
At a step $t=k_0 + m(N-1)$ ($k_0=1, \cdots, N-1$ and $m=0,1, \cdots$), we randomly choose $j \in \{i_1, \cdots, i_{N-1}  \}$.
If $j \neq i_{k_0}$, do nothing. If $j =i_{k_0}$ and $x_j = y_j$, flip both of them.
If $j =i_{k_0}$ and $x_j \neq y_j$, flip $x_j$ and $y_{\alpha(j)}$ where $\alpha(j)$ is the position satisfying
$x_{\alpha(j)} \neq y_{\alpha(j)}$ next to $j$. For instance, when $X=(0,1,0,1,1)$ and $Y=(1,1,0,0,0)$, $\alpha(1)=4$ and $\alpha(4)=5$.
Each of $X_k$ and $Y_k$ is equivalent to the Markov chain $\tilde{\mathcal{M}}$ and
they also satisfy $X_{k+1} = Y_{k+1}$ if $X_{k} = Y_{k}$. Hence, $(X_k, Y_k)$ is a coupling of $\tilde{\mathcal{M}}$.

To investigate the coupling, we use the property of a special type of Markov chain, a {\it coupon collecting}.
A coupon collecting of $r$-coupons is a Markov chain on a set of states $\{0, 1, \cdots, r \}$ with transition matrix given by
\begin{align*}
{\rm Prob} [X_{t+1} = k+1 |X_t = k ] &= \frac{r-k}{r}, \\
{\rm Prob} [X_{t+1} = k |X_t = k ] &= \frac{k}{r}. 
\end{align*}
A coupon collecting of $r$ coupons is interpreted as a trial of collecting a complete set of $r$ different coupons by drawing one coupon at each step. When $r_0$ coupons are initially at hand, we denote the necessary number of steps to draw all coupons by $\tau_{\rm coupon}^{(r,r_0)}$.

\begin{Proposition} \label{Prop:Mar1}
{\it Let $(X_k, Y_k)$ be a coupling of $\tilde{\mathcal{M}}$ defined above and $T_{xy}$ be its stopping time. Then $T_{xy}$ is bounded from above by
\begin{equation}
T_{xy} \leq (N-1) \tau_{\rm coupon}^{(N-1,N-2)}.  
\end{equation}}
\end{Proposition}

\begin{Proof}
By definition of the coupling,
for a fixed $k_0 \in \{1,\cdots, N-1 \}$, once $i_{k_0}$ is chosen at the steps $k=k_0+  m(N-1)$ ($m=0,1,\cdots$), we have $x_{i_{k_0}}=y_{i_{k_0}}$.
The number of steps necessary for picking up $k_0$ from $\{1,\cdots, N-1 \}$ is equal to that of
a coupon collecting of $(N-1)$-coupons with initially $N-2$ coupons at hand, that is, $\tau_{\rm coupon}^{(N-1,N-2)}$.
Since there exists $N-1$ choices of $k_0$, the stopping time $T_{xy}$ is smaller than or equal to $(N-1)  \tau_{\rm coupon}^{(N-1,N-2)}$.
\begin{flushright}$\blacksquare$\end{flushright}
\end{Proof}

A coupon collecting of $r$-coupons starting with no coupon is a well-studied problem. 
If we draw coupons $r \log r $ times, we can collect all $r$ coupons with high probability, that is,
$\tau_{\rm coupon}^{(r,0)}$ is typically given by $r \log r$.
For $\tau_{\rm coupon}^{(r,r_0)}$, we first show that
\begin{equation}
{\rm Prob} [ \tau_{\rm coupon}^{(r,r_0)} > r (\log(r-r_0) + c )] \leq e^{-c}, \label{Eq:Mar2}
\end{equation}
for any $c>0$. This is shown in a standard way (see~\ref{Ap:Mar} for details).

By appealing to Theorem~\ref{Thm:Stop}, Proposition~\ref{Prop:Mar1} and Eq.~\eqref{Eq:Mar2}, we can bound
$\Delta((N-1)^2 c )$ by
\begin{align*}
\Delta((N-1)^2 c ) &\leq \max_{x,y} {\rm Prob}[T_{xy} > (N-1)^2 c) ]   \\
&\leq {\rm Prob}[ (N-1) \tau_{\rm coupon}^{(N-1,N-2)} > (N-1)^2 c ]  \\
&={\rm Prob}[  \tau_{\rm coupon}^{(N-1,N-2)} > (N-1) c ]   \\
&<e^{-c}. 
\end{align*}
Denote by $\tilde{T}_{\rm mix}(\epsilon')$ the mixing time of the Markov chain $\tilde{\mathcal{M}}$ with an error $\epsilon'$.
Since $\tilde{T}_{\rm mix}(\epsilon')$ is defined by $\Delta(\tilde{T}_{\rm mix}(\epsilon')) \leq \epsilon'$, we obtain
\begin{equation*}
\tilde{T}_{\rm mix}(\epsilon') \leq (N-1)^2  \log \epsilon'^{-1},  
\end{equation*}
which also provides an upper bound on the mixing time of the Markov chain $\mathcal{M}$ since $T_{\rm mix}(\epsilon') < \tilde{T}_{\rm mix}(\epsilon')$.

\subsubsection{Upper and lower bounds of the convergence time $T_{\rm conv}(\epsilon)$} \label{SSS:ULbounds}

We require an error $\epsilon/(\Pi_{\rm sum} 2^{7N})$ in order to obtain an upper bound on the convergence time $T_{\rm conv}(\epsilon)$ (see Eq.~\eqref{Eq:CoeffError}) and, an error $\epsilon/\Pi_{\rm sum}$ for a lower bound (see Eq.~\eqref{Eq:CoeffErrorLower}).
Recalling that the unitary operation $\tilde{W}$ consists of $T_{\tilde{W}} = \lceil N/2 \rceil$ two-qubit gates,
we obtain bounds for $T_{\rm conv}(\epsilon)$ such that
\begin{multline*}
(\frac{N(N-1)}{4} -1)  \log (\frac{\Pi_{\rm sum}(L^{(\pm)}, R^{(\pm)}) }{2\epsilon}) \leq
T_{\rm conv}(\epsilon)- T_{\tilde{W}} \\
 \leq  (N-1)^2 \log (\frac{2^{7 N} \Pi_{\rm sum}(L^{(\pm)}, R^{(\pm)}) }{\epsilon}).  
\end{multline*}

It is straightforward to show that $\Pi_{\rm sum}(L^{(\pm)}, R^{(\pm)}) \leq 1$ for any $(L^{(\pm)}, R^{(\pm)})$ and any $\rho_0$.
For the lower bound, there exist $\rho_0$ and $(L^{(\pm)}, R^{(\pm)})$ such that $\Pi_{\rm sum}(L^{(\pm)}, R^{(\pm)}) = 1$.
Therefore, we finally obtain
\begin{equation*}
\frac{N}{2}  + \biggl(\frac{N^2}{4} + O(N) \biggr) \log (2\epsilon)^{-1}    \leq T_{\rm conv}(\epsilon) 
\leq 7 N^3 \log2 + N^2 \log\epsilon^{-1} + O(N^2),  
\end{equation*}
for any initial state $\rho_0$.
This concludes the proof of Theorem~\ref{Thm:phase-random circuitCZ}.

\section{Summary and concluding remarks} \label{Sec:Sum}

In this paper, we have introduced concepts of diagonal-unitary $t$-designs and toric $t$-designs 
that simulate up to the $t$-th order of statistical moments of diagonal-unitary matrices and phase-random states, respectively.   
We have presented how to implement diagonal-unitary $2$-designs with the computational basis for $N$-qubit systems by using two types of the phase-random circuits,  the CP and CZ phase-random circuit.  
We have shown that the CP phase-random circuit exactly achieves a diagonal-unitary $2$-design after applying two-qubit diagonal gates on 
all pairs of qubits, which requires $N(N-1)/2$ gates.
On the other hand, the CZ phase-random circuit approximately achieves a diagonal-unitary $2$-design after applying at most $O(N^2(N+\log1/\epsilon))$ two-qubit diagonal gates
on randomly chosen pairs of qubits. Due to the random choice of pairs, the number of gates exceeds $N(N-1)/2$ despite the commutativity of gates.
Our results show that random variables in the genuine two-qubit diagonal gate provide a stronger ability for randomizing phases.  
We have also presented a protocol generating an exact complex-projective $2$-design by using the CP phase-random circuit, which
is more feasible to implement in experiments comparing to previously known protocols.

In analogy with random circuits, which are shown to approximately achieve unitary $t$-designs for any $t$ by applying $poly(N,t)$ two-qubit gates\cite{BHH2012},  it is natural to expect that the phase-random circuits with appropriate gate sets would also achieve diagonal-unitary $t$-designs in $poly(N,t)$ iterations.  
However, this is not the case as long as we use two-qubit diagonal gates since 
there is a lack of the number of parameters due to the commutativity of gates.
Thus, the gate set should include multi-qubit gates if only diagonal gates are used for constructing diagonal-unitary $t$-designs for large $t$.
It is interesting to specify the diagonal gate set of the phase-random circuit achieving diagonal-unitary $t$-designs for arbitrary $t$
and to construct a quantum circuit composed of non-diagonal two-qubit gates that achieves diagonal-unitary $t$-designs,
which will be addressed in a separate paper\cite{Prep}.

\section*{Acknowledgment}
The authors thank F. G. S. L. Brandao for insightful comments on unitary $t$-designs, 
M. Koashi for important suggestions for the CP phase-random circuit, and
P. S. Turner for helpful discussions.   This work was supported by Project for Developing Innovation Systems of the Ministry of Education, Culture, Sports, Science and Technology (MEXT), Japan. Y.~N. acknowledges support from JSPS by KAKENHI (Grant No. 222812) and M.~M. acknowledges support from JSPS by KAKENHI  (Grant No. 23540463).

\appendix

\section{Discrete phases are sufficient to achieve diagonal-unitary $t$-designs} \label{Ap:DiscreteP}

We show that a set of unitary matrices $\Omega_t=\{ \sum_n e^{i \phi_n} \ketbra{u_n}{u_n} \}$,
where $\phi_n$ is randomly chosen from $\{\frac{2 \pi k}{t+1} \}_{k=0,1,\cdots, t}$, is a diagonal-unitary $t$-design in the basis $\{\ket{u_n}\}$.

For $U_{\phi} = \sum_n e^{i \phi_n} \ketbra{u_n}{u_n}$, $U_{\phi}^{\otimes t} \otimes U_{\phi}^{{\dagger} \otimes t}$ is calculated to
\begin{equation}
U_{\phi}^{\otimes t} \otimes U_{\phi}^{{\dagger} \otimes t} = \sum_{n_k, m_k}
 \exp[i \sum_{k=1}^t (\phi_{n_k} - \phi_{m_k})] 
\ketbra{u_{n_1} \cdots u_{n_t} u_{m_1} \cdots u_{m_t}}{u_{n_1} \cdots u_{n_t}u_{m_1} \cdots u_{m_t}}, \notag
\end{equation}
where the summations are taken over $n_1,\cdots, n_t=1 ,\cdots, d$ and $m_1,\cdots, m_t=1 ,\cdots, d$.

For $\mathcal{U}_{\rm diag}^{(t)}$ and $\Omega_t$, the expectation of an operator $X$ is taken over $\phi_i \in [0,2 \pi)$ and $\phi_i^{\prime} \in \{\frac{2 \pi k}{t+1} \}_{k=0,1,\cdots, t}$ for all $i=1, \cdots, d$, respectively, namely,
\begin{align*}
&\mathbb{E}_{\mathcal{U}_{\rm diag}^{(t)}}[X] =\frac{1}{(2 \pi)^d}\int_0^{2\pi} X d\phi_1 \cdots d\phi_d,\\
&\mathbb{E}_{\Omega_t}[X] =( \frac{1}{t+1})^d \sum_{\phi_1^{\prime} = 0, \cdots, \frac{2 \pi t}{t+1}}  \cdots \sum_{\phi_d^{\prime} = 0, \cdots, \frac{2 \pi t}{t+1}}  X.
\end{align*}
The equation $\mathbb{E}_{\Omega_t}[U^{\otimes t} \otimes U^{{\dagger} \otimes t}]=\mathbb{E}_{\mathcal{U}_{\rm diag}^{(t)}}[U^{\otimes t} \otimes U^{{\dagger} \otimes t}]$ follows from an identity that
\begin{multline}
\frac{1}{(2 \pi)^d}\int_0^{2\pi} \exp[i \sum_{k=1}^t ( \phi_{n_k} - \phi_{m_k})]d\phi_1 \cdots d\phi_d =\\
 ( \frac{1}{t+1})^d \sum_{\phi_1^{\prime} = 0, \cdots, \frac{2 \pi t}{t+1}}  \cdots \sum_{\phi_d^{\prime} = 0, \cdots, \frac{2 \pi t}{t+1}} \exp[i \sum_{k=1}^t (\phi_{n_k}^{\prime} - \phi_{m_k}^{ \prime} )]. \notag
\end{multline}
Thus, discrete phases are sufficient to achieve a diagonal-unitary $t$-design.

\section{Discrete phases in the $CP$ phase-random circuits} \label{App:DiscretePhasesCP}

We show that, in the CP phase-random circuits,
it is sufficient to choose the phases $\alpha$ and $\beta$ from $\{0, \frac{2 \pi}{3}, \frac{4 \pi}{3}\}$, 
and $\gamma$ from $\{0, \pi\}$, instead of choosing the phases uniformly from $[0, 2\pi)$.

To show this, we consider 
two expectations of $W:=(\prod_{i < j} W_{ij}(\alpha_{ij}, \beta_{ij}, \gamma_{ij}))^{\otimes 2} \otimes (\prod_{i < j} W_{ij}^{\dagger}(\alpha_{ij}, \beta_{ij}, \gamma_{ij}))^{\otimes 2}$,
where $W_{ij}(\alpha_{ij}, \beta_{ij}, \gamma_{ij}) \in \mathcal{W}_{\rm diag}^{CP}$ acts on the $i$th and $j$th qubits with phases $\alpha_{ij}, \beta_{ij}, \gamma_{ij}$.
One expectation is taken over $\alpha, \beta, \gamma \in [0, 2\pi)$ and is denoted by $\mathbb{E}_{\rm cont}[W]$.
The other is taken over $\alpha, \beta \in \{0, \frac{2 \pi}{3}, \frac{4 \pi}{3}\}$ and $\gamma \in \{0, \pi\}$ and is denoted by $\mathbb{E}_{\rm disc}[W]$.
Due to the fact that all $\alpha_{ij}, \beta_{ij}$ and $\gamma_{ij}$ are independent, both expectations are given by
\begin{equation*}
\mathbb{E}[W] = \prod_{i < j} \mathbb{E}[W_{ij}(\alpha_{ij}, \beta_{ij}, \gamma_{ij})^{\otimes 2} \otimes W_{ij}^{\dagger}(\alpha_{ij}, \beta_{ij}, \gamma_{ij})^{\otimes 2}].
\end{equation*}
In $W_{ij}(\alpha_{ij}, \beta_{ij}, \gamma_{ij})^{\otimes 2} \otimes W_{ij}^{\dagger}(\alpha_{ij}, \beta_{ij}, \gamma_{ij})^{\otimes 2}$, 
all terms except those equal to $1$, e.g., $e^{i \alpha_{ij}}$, $e^{i \beta_{ij}}$, $e^{i 2\alpha_{ij}}$ and $e^{i ( \alpha_{ij} + \beta_{ij} + \gamma_{ij})}$, 
disappear by integrating over $\alpha_{ij}, \beta_{ij}$ and $\gamma_{ij}$.
This is the case even if we average over $\alpha_{ij}, \beta_{ij} \in \{0, \frac{2 \pi}{3}, \frac{4 \pi}{3}\}$ and $\gamma_{ij} \in \{0, \pi\}$,
since $W_{ij}^{\otimes 2}(\alpha_{ij}, \beta_{ij}, \gamma_{ij}) \otimes W_{ij}^{\dagger}(\alpha_{ij}, \beta_{ij}, \gamma_{ij})^{\otimes 2}$ does not contain terms such as $e^{3i \alpha_{ij}}$, $e^{3i \beta_{ij}}$ and
$e^{2 i \gamma}$.
Thus, it is shown that $\mathbb{E}_{\rm cont}[W] = \mathbb{E}_{\rm disc}[W]$, namely, 
$\alpha_{ij}, \beta_{ij}$ can be chosen from $\{0, \frac{2 \pi}{3}, \frac{4 \pi}{3}\}$ and $\gamma_{ij}$ from $\{0, \pi\}$.

\section{Necessary and sufficient conditions for an $\epsilon$-approximate diagonal-unitary $2$-design in terms of 
the expansion coefficients} \label{App:NecSuff}

An $\epsilon$-approximate diagonal-unitary $2$-design $\mathcal{U}^{(2,\epsilon)}_{\rm diag}$ is defined by
\begin{equation}
|\!| \mathcal{E}^{(2)}_{\mathcal{U}^{(2,\epsilon)}_{\rm diag}} - \mathcal{E}^{(2)}_{\mathcal{U}_{\rm diag}} |\!|_{\diamond} \leq \epsilon,
\label{approxCZ}
\end{equation}
where $\mathcal{U}_{\rm diag}$ is random diagonal-unitary matrices and $\mathcal{E}^{(2)}_{\mathcal{V}}$ for an ensemble of unitary matrices $\mathcal{V}$ is a superoperator defined by Eq.~\eqref{Eq:superop}.
We show that sufficient and necessary conditions for Eq.~\eqref{approxCZ} to hold in terms of $ \mathbb{E}_{C_T^{CZ}}[ \xi_T (\mathbf{p},\mathbf{q})] $ are given by
\begin{equation}
\forall (\mathbf{p},\mathbf{q}), \biggl| \mathbb{E}_{C_T^{CZ}}[ \xi_T (\mathbf{p},\mathbf{q})  ]-\mathbb{E}_{C_{\infty}^{CZ}}[ \xi_{\infty} (\mathbf{p},\mathbf{q})  ] \biggr| \leq \frac{\epsilon}{2^{7N}}, \label{Eq:111111111}
\end{equation}
and
\begin{equation}
\forall (\mathbf{p},\mathbf{q}), \biggl| \mathbb{E}_{C_T^{CZ}}[ \xi_T (\mathbf{p},\mathbf{q})  ]-\mathbb{E}_{C_{\infty}^{CZ}}[ \xi_{\infty} (\mathbf{p},\mathbf{q})  ] \biggr| \leq \epsilon,
\end{equation}
respectively, for any initial state $\rho_0$ on a $2N$-qubit system.

First, we assume Eq.~\eqref{Eq:111111111}. It implies
\begin{align}
&\forall \rho_0, \sum_{(\mathbf{p}, \mathbf{q})}  \biggl| \mathbb{E}_{C_T^{CZ}}[ \xi_T (\mathbf{p},\mathbf{q})  ]-\mathbb{E}_{C_{\infty}^{CZ}}[ \xi_{\infty} (\mathbf{p},\mathbf{q})  ] \biggr| \leq  \frac{\epsilon}{2^{3N}} \\
&\Rightarrow 
\forall \rho_0, \biggl|\! \biggl| \sum_{(\mathbf{p}, \mathbf{q})}  \biggr( \mathbb{E}_{C_T^{CZ}}[ \xi_T (\mathbf{p},\mathbf{q})  ]-\mathbb{E}_{C_{\infty}^{CZ}}[ \xi_{\infty} (\mathbf{p},\mathbf{q})  ] \biggr) \sigma_{\mathbf{p}} \otimes \sigma_{\mathbf{q}} \biggr|\! \biggr|_1 \leq  \frac{\epsilon}{2^{N}}\\
&\Leftrightarrow \forall \rho_0,   |\!| \mathcal{E}^{(2)}_{\mathcal{U}^{(2,\epsilon)}}(\rho_0) - \mathcal{E}^{(2)}_{\mathcal{U}}(\rho_0) |\!|_{1} \leq \frac{\epsilon}{ 2^{2N}}\\
&\Rightarrow |\!| \mathcal{E}^{(2)}_{\mathcal{U}^{(2,\epsilon)}} - \mathcal{E}^{(2)}_{\mathcal{U}} |\!|_{\diamond} \leq \epsilon,
\end{align}
where the Cauchy-Schwarz inequality is used to obtained the first relation and, a relation that 
$\forall \rho, | \! | \mathcal{E}(\rho) |\! |_1 \leq \alpha$ implies 
$| \! | \mathcal{E} |\! |_{\diamond} \leq d \alpha$, where $d$ is a dimension of the space that $\rho$ acts on, is used to obtain the last relation.
This provides the sufficient condition.

To obtain a necessary condition, we start from 
$|\!| \mathcal{E}^{(2)}_{\mathcal{U}^{(2,\epsilon)}} - \mathcal{E}^{(2)}_{\mathcal{U}} |\!|_{\diamond} \leq \epsilon$ and use a fact that
$| \! | \mathcal{E} |\! |_{\diamond} \geq | \! | \mathcal{E}(\rho) |\! |_1$ for any state $\rho$.
In terms of the expansion coefficients, we obtain
\begin{align}
&\forall \rho_0, 2^{-N} \biggl|\! \biggl| \sum_{(\mathbf{p}, \mathbf{q})}  \biggr( \mathbb{E}_{C_T^{CZ}}[ \xi_T (\mathbf{p},\mathbf{q})  ]-\mathbb{E}_{C_{\infty}^{CZ}}[ \xi_{\infty} (\mathbf{p},\mathbf{q})  ] \biggr) \sigma_{\mathbf{p}} \otimes \sigma_{\mathbf{q}} \biggr|\! \biggr|_1 \leq \epsilon\\
&\Rightarrow  \forall \rho_0, 2^{-N} \biggl|\! \biggl| \sum_{(\mathbf{p}, \mathbf{q})}  \biggr( \mathbb{E}_{C_T^{CZ}}[ \xi_T (\mathbf{p},\mathbf{q})  ]-\mathbb{E}_{C_{\infty}^{CZ}}[ \xi_{\infty} (\mathbf{p},\mathbf{q})  ] \biggr) \sigma_{\mathbf{p}} \otimes \sigma_{\mathbf{q}} \biggr|\! \biggr|_2 \leq \epsilon\\
&\Leftrightarrow  \forall \rho_0, 
\sum_{(\mathbf{p}, \mathbf{q})}  \biggl( \mathbb{E}_{C_T^{CZ}}[ \xi_T (\mathbf{p},\mathbf{q})  ]-\mathbb{E}_{C_{\infty}^{CZ}}[ \xi_{\infty} (\mathbf{p},\mathbf{q})  ] \biggr)^2 \leq  \epsilon^2 \\
&\Rightarrow  \forall \rho_0, \forall (\mathbf{p},\mathbf{q}), \biggl| \mathbb{E}_{C_T^{CZ}}[ \xi_T (\mathbf{p},\mathbf{q})  ]-\mathbb{E}_{C_{\infty}^{CZ}}[ \xi_{\infty} (\mathbf{p},\mathbf{q})  ] \biggr| \leq \epsilon,
\end{align}
where $|\!| X |\! |_2 = \sqrt{\tr X^{\dagger} X}$ is a Hilbert-Schmidt norm.
Thus we obtain the necessary condition.


\section{Calculation of $G_{ij}(\mathbf{p},\mathbf{q};\mathbf{p}',\mathbf{q}')$} \label{Ap:G}

$G_{ij}(\mathbf{p},\mathbf{q};\mathbf{p}',\mathbf{q}')$ is defined by
\begin{equation}
G_{ij}(\mathbf{p},\mathbf{q};\mathbf{p}',\mathbf{q}')=\mathbb{E}[ \tr \sigma_{\mathbf{p}} W_{ij}\sigma_{\mathbf{p}'} W_{ij}^{\dagger} \tr \sigma_{\mathbf{q}} W_{ij}\sigma_{\mathbf{q}'} W_{ij}^{\dagger}], \notag
\end{equation}
where $W_{ij}$ is a two-qubit diagonal gate on the $i$-th and $j$-th qubits and randomly chosen from the gate set 
$\mathcal{W}^{CZ}=\{ diag(1, e^{i \alpha}, e^{i \beta}, -e^{i(\alpha+\beta)}) \}_{\alpha,\beta}$ or $\mathcal{W}^{CP}=\{ diag(1, e^{i \alpha}, e^{i \beta}, e^{i\gamma}) \}_{\alpha,\beta,\gamma}$.
Since $\mathcal{W}^{CP}$ is more general than $\mathcal{W}^{CZ}$, we start with the calculation of
$G_{ij}(\mathbf{p},\mathbf{q};\mathbf{p}',\mathbf{q}')$ for $\mathcal{W}^{CP}$.

In order to calculate $G_{ij}(\mathbf{p},\mathbf{q};\mathbf{p}',\mathbf{q}')$, we define $\mathcal{D}_{ab}, \mathcal{E}_{ab}$ and $\Delta_a$ by
\begin{align*}
\mathcal{D}_{ab} &= \delta_{a0}\delta_{bz} + \delta_{az}\delta_{b0},\\
\mathcal{E}_{ab} &= \delta_{ax}\delta_{by} - \delta_{ay}\delta_{bx},
\end{align*}
and
\begin{equation}
\Delta_a = \delta_{a0} + \delta_{az} - \delta_{ax} - \delta_{ay}. \notag
\end{equation}
We also use a notation that $\delta_{n \in S} = 1$ if $n \in S$ and $\delta_{n \in S} = 0$ if $n \notin S$.

For $W_{ij}=diag(1, e^{i \alpha}, e^{i \beta},e^{i \gamma})$, it is straightforward to calculate $\tr \sigma_{\mathbf{p}} W_{ij}\sigma_{\mathbf{p}'} W_{ij}^{\dagger}$  and we obtain
\begin{align*}
&\frac{1}{2^N}\tr \sigma_{\mathbf{p}} W_{ij}\sigma_{\mathbf{p}'} W_{ij}^{\dagger} \notag \\
&=
\delta_{\mathbf{p}, \mathbf{p}'}
\biggl\{ \delta_{p_i, p_j \in \{0,z \}} + \frac{1}{2} \delta_{p_i \in \{x,y \}} \bigl(\cos \beta + \cos ( \alpha-\gamma)\bigr) 
+\frac{1}{2} \delta_{p_j \in \{x,y \}} \bigl(\cos \alpha + \cos ( \beta-\gamma)\bigr) \notag \\
&\hspace{10mm}-\frac{1}{2} \delta_{p_i, p_j \in \{x,y \}} \bigl[\cos \alpha + \cos \beta -\cos \gamma - \cos ( \alpha-\beta) + \cos ( \alpha-\gamma) + \cos ( \beta-\gamma) \bigr] \biggr\} \notag \\
&+
\frac{1}{2} \delta_{p_i,p_i'} \biggl\{ \delta_{p_i \in \{x,y \}}
 \bigl(\cos \beta - \cos ( \alpha-\gamma)\bigr) \mathcal{D}_{p_j, p_j'} \notag \\
&\hspace{10mm} -\bigl[\delta_{p_i \in \{0,z \}} \bigl(\sin \alpha - \sin (\beta - \gamma)\bigr)
+  \delta_{p_i \in \{x,y \}} \bigl(\sin \gamma + \sin (\alpha - \beta)\bigr) \bigr] \mathcal{E}_{p_j, p_j'} \biggr\}\notag \\
&+
\frac{1}{2} \delta_{p_j,p_j'} \biggl\{ \delta_{p_j \in \{x,y \}}
 \bigl(\cos \alpha - \cos ( \beta-\gamma)\bigr) \mathcal{D}_{p_i, p_i'} \notag \\
& \hspace{10mm}-\bigl[\delta_{p_j \in \{0,z \}} \bigl(\sin \beta - \sin (\alpha - \gamma)\bigr) 
+  \delta_{p_j \in \{x,y \}} \bigl(\sin \gamma + \sin (\beta - \alpha)\bigr) \bigr] \mathcal{E}_{p_i, p_i'} \biggr\} \notag \\
&-\frac{1}{2}\bigl(\sin \beta + \sin(\alpha-\gamma) \bigr) \mathcal{E}_{p_i, p_i'} \mathcal{D}_{p_j, p_j'}
-\frac{1}{2}\bigl(\sin \alpha + \sin(\beta-\gamma)\bigr) \mathcal{D}_{p_i, p_i'} \mathcal{E}_{p_j, p_j'} \notag \\
&\hspace{10mm} +\frac{1}{2}\bigl(\cos(\alpha- \beta) - \cos \gamma \bigr) \mathcal{E}_{p_i, p_i'} \mathcal{E}_{p_j, p_j'}. \label{Eq:Gen}
\end{align*} 

In the case of $\mathcal{W}^{CP}$, by taking the average over $\alpha, \beta=0, \frac{2 \pi}{3},\frac{4 \pi}{3}$ and $\gamma=0, \pi$,
$G_{ij}$ is calculated to
\begin{align*}
\frac{1}{2^{2N}}G_{ij}(\mathbf{p},\mathbf{q};\mathbf{p}',\mathbf{q}')
=&
\delta_{\mathbf{p},\mathbf{p}'}\delta_{\mathbf{q},\mathbf{q}'}
\biggl(\delta_{p_i,p_j,q_i,q_j \in \{0,z \}}+\frac{1}{4} \delta_{p_i,q_i \in \{0,z \}} \delta_{p_j,q_j \in \{x,y \}}  \\
&+\frac{1}{4}\delta_{p_i,q_i \in \{x,y \}} \delta_{p_j,q_j \in \{0,z \}} +  \frac{1}{4}\delta_{p_i,p_j,q_i,q_j \in \{x,y \}} \biggr) \\
&+
\frac{1}{4} \delta_{p_i,q_i \in \{x,y \}}\delta_{p_i,p_i'}\delta_{q_i,q_i'}
\biggl( \mathcal{D}_{p_j, p_j'}\mathcal{D}_{q_j, q_j'} + \mathcal{E}_{p_j, p_j'}\mathcal{E}_{q_j, q_j'}   \biggr)\\
&+
\frac{1}{4} \delta_{p_j,q_j \in \{x,y \}}\delta_{p_j,p_j'}\delta_{q_j,q_j'}
\biggl( \mathcal{D}_{p_i, p_i'}\mathcal{D}_{q_i, q_i'} + \mathcal{E}_{p_i, p_i'}\mathcal{E}_{q_i, q_i'}   \biggr)\\
&+
\frac{1}{4} \mathcal{E}_{p_i, p_i'}\mathcal{E}_{q_i, q_i'}
\biggl(\delta_{p_j,q_j \in \{0,z \}}\delta_{p_j,p_j'}\delta_{q_j,q_j'} +  \mathcal{D}_{p_j, p_j'}\mathcal{D}_{q_j, q_j'}  \biggr)\\
&+
\frac{1}{4} \mathcal{E}_{p_j, p_j'}\mathcal{E}_{q_j, q_j'}
\biggl(\delta_{p_i,q_i \in \{0,z \}}\delta_{p_i,p_i'}\delta_{q_i,q_i'} +  \mathcal{D}_{p_i, p_i'}\mathcal{D}_{q_i, q_i'}  \biggr)\\
&+\frac{1}{4}\mathcal{E}_{p_i, p_i'}\mathcal{E}_{q_i, q_i'}\mathcal{E}_{p_j, p_j'}\mathcal{E}_{q_j, q_j'}.
\end{align*}

By investigating each case, we obtain Eq.~\eqref{Eq:EvoCP}.

On the other hand, in the case of $\mathcal{W}^{CZ}$, $\gamma$ is set to be $\alpha+\beta + \pi$ and
$G_{ij}$ is obtained as
\begin{align*}
\frac{1}{2^{2N}}G_{ij}(\mathbf{p},\mathbf{q};\mathbf{p}',\mathbf{q}')
=&
\delta_{\mathbf{p},\mathbf{p}'}\delta_{\mathbf{q},\mathbf{q}'}\biggl(\delta_{p_i,p_j,q_i,q_j \in \{0,z \}}+\frac{1}{4}\delta_{p_i,p_j,q_i,q_j \in \{x,y \}} \biggr) \\
&+\frac{1}{2} \delta_{p_i,q_i \in \{x,y \}}\delta_{p_i,p_i'}\delta_{q_i,q_i'}\biggl( \mathcal{D}_{p_j, p_j'}\mathcal{D}_{q_j, q_j'}+\frac{1}{2}\mathcal{E}_{p_j, p_j'}\mathcal{E}_{q_j, q_j'}   \biggr)\\
&+\frac{1}{2} \delta_{p_j,q_j \in \{x,y \}}\delta_{p_j,p_j'}\delta_{q_j,q_j'}\biggl( \mathcal{D}_{p_i, p_i'}\mathcal{D}_{q_i, q_i'}+\frac{1}{2}\mathcal{E}_{p_i, p_i'}\mathcal{E}_{q_i, q_i'}   \biggr)\\
&+\frac{1}{2}\mathcal{D}_{p_i, p_i'}\mathcal{D}_{q_i, q_i'}\mathcal{E}_{p_j, p_j'}\mathcal{E}_{q_j, q_j'} 
+\frac{1}{2}\mathcal{E}_{p_i, p_i'}\mathcal{E}_{q_i, q_i'}\mathcal{D}_{p_j, p_j'}\mathcal{D}_{q_j, q_j'} \\
&+\frac{1}{4}\mathcal{E}_{p_i, p_i'}\mathcal{E}_{q_i, q_i'}\mathcal{E}_{p_j, p_j'}\mathcal{E}_{q_j, q_j'} 
\end{align*}
leading to Eq.~(\ref{Eq:Evo1}).

\section{Introduction of Markov chain} \label{Ap:Markov}

A Markov chain is a sequence of random variables indexed by a discrete {\it step} $t \in \mathbb{N}$ that take values in a set of {\it states} $S=\{ s \}$.  
We define a probability distribution $\{ \Pi_t(s)\}_{s \in S}$ at a step $t$ over the state space $S$.  
The Markov property is that the probability distribution $\Pi_{t+1}$ depends only on $\Pi_t$.
This evolution of the probability distribution is governed by a stochastic {\it transition matrix} $\mathcal{P}$ 
such that $\Pi_{t+1}= \mathcal{P} \Pi_{t}$.
The elements of a transition matrix are denoted by $\mathcal{P}(s, s')$, which represents the probability that a transition from $s$ to $s'$ occurs.
Using an initial distribution $\Pi_0$, the probability distribution at step $t$ is given by $\Pi_t = \mathcal{P}^t \Pi_0$.
A Markov chain is said to be irreducible (aperiodic) when the transition matrix is irreducible (aperiodic).
For an irreducible and aperiodic Markov chain, the Perron-Frobenius theorem guarantees that there exists a unique 
stationary distribution $\Pi_\infty=\lim_{t \rightarrow \infty} \Pi_t$ independent of the initial probability distribution.

We define the {\it mixing time}.
The mixing time is the number of steps required for the actual distribution to be close to the stationary distribution, 
where the distance after $t$-steps is defined by
\begin{equation*}
\Delta (t):=\max_{s  \in S} |\Pi_t(s) - \Pi_\infty(s)|.  
\end{equation*}
We define the mixing time $T_{\mathrm{mix}}(\epsilon')$ such that for any $\epsilon' >0$ 
\begin{equation*}
T_{\mathrm{mix}}(\epsilon') := \min\{t | \Delta (t') \leq \epsilon' \text{ \ for all \ } t' \geq t \}.  
\end{equation*}

In order to study an upper bound and a lower bound on the mixing time, we introduce the {\it relaxation time} of a Markov chain.
Denote the eigenvalues of a transition matrix $\mathcal{P}$ by $\lambda_i$ ($i=1,2,\cdots$) in decreasing order.
When a transition matrix is irreducible and aperiodic, $1=\lambda_1>\lambda_2$. 
The {\it relaxation time} $T_{\rm rel}$ is defined by
\begin{equation*}
T_{\rm rel} = (1- \max_{i=2,3,\cdots}|\lambda_i| )^{-1},  
\end{equation*}
which gives bounds of the mixing time $T_{\mathrm{mix}}(\epsilon')$ such that
\begin{equation}
(T_{\rm rel}-1) \log(\frac{1}{2\epsilon'}) \leq  T_{\mathrm{mix}}(\epsilon') \leq \log(\frac{1}{\epsilon' \Pi_{\rm min}}) T_{\rm rel}, \label{Eq:RelaxMarkov}
\end{equation}
where $\Pi_{\rm min}:=\min_s \Pi_{\infty}(s) $ is the minimum stationary probability \cite{LPW2009}.

Although the relaxation time provides both of the upper and the lower bound on the mixing time,
it does not give tight bounds. Hence, we introduce a {\it coupling} method for investigating the upper bound on the mixing time.
A pair of two random walks $(X_t, Y_t)$, where $t$ denotes the number of steps, is said to be a coupling of a Markov chain when
the following two conditions are satisfied. First, $X_t$ and
$Y_t$ is each a faithful copy of the Markov chain.
Second, $(X_t, Y_t)$ should satisfy the condition that $X_t = Y_t$ implies $X_{t+1} = Y_{t+1}$.
For a coupling $(X_t, Y_t)$, we define the {\it stopping time} $T_{xy}$ by
\begin{equation*}
T_{xy} := \min\{t | X_t=Y_t, \text{ when } X_0=x, Y_0=y  \}. \nonumber 
\end{equation*}
By definition, $X_t=Y_t$ for all $t > T_{xy}$. The stopping time is related to the mixing time through the following 
theorem\cite{LPW2009}.

\begin{Theorem} \label{Thm:Stop}
{\it Let $(X_t, Y_t)$ be a coupling of a Markov chain and $T_{xy}$ be the stopping time. Then,
\begin{equation*}
\Delta(t) \leq \max_{x,y} {\rm Prob}[T_{xy} > t].  
\end{equation*}}
\end{Theorem}
Since the mixing time is obtained from $\Delta(t)$, we can derive an upper bound on the mixing time from the stopping time.

In the main text, we use a relaxation time to obtain 
a lower bound on the mixing time and investigate an upper bound on the mixing time by using the coupling method.

\section{Coupon collecting starting with non-zero coupons} \label{Ap:Mar}
We consider a coupon collecting and show that for $c>0$,
\begin{equation}
{\rm Prob} [ \tau_{\rm coupon}^{(r,r_0)} > r (\log(r-r_0) + c )] \leq e^{-c}.
\label{eqn:couponprob}
\end{equation}

Suppose that any of the $j$-th coupons ($j=1, \cdots, r-r_0$) are initially not at hand.
We denote by $A_j$ an event where the $j$-th coupon has not been collected within $r (\log(r-r_0) + c)$ steps.
An upper bound of the left hand side of Eq.~(\ref{eqn:couponprob}) is obtained by
\begin{align*}
{\rm Prob} [ \tau_{\rm coupon}^{(r,r_0)} > r (\log(r-r_0) + c )]
&={\rm Prob} [ \cup_{j=1}^{r-r_0} A_{j}]\\
&\leq \sum_{j=1}^{r-r_0} {\rm Prob} [ A_{j}]\\
&=\sum_{j=1}^{r-r_0} (1-\frac{1}{r})^{r (\log(r-r_0) + c)}\\
&=(r-r_0) (1-\frac{1}{r})^{r (\log(r-r_0) + c)}\\
&\leq (r-r_0) \exp[- \log(r-r_0) -c ]\\
&= e^{-c}. 
\end{align*}

\end{document}